\documentclass[11pt,a4paper]{article}
\usepackage{jheppub}
\usepackage{subfigure}
\usepackage{cases}
\usepackage{multirow}

\title{Stability analysis for $Q$-balls with spectral method}

\author[a, b]{Qian Chen,}

\author[b,c]{Lars Andersson,}

\author[a,d,e]{Li Li}

\affiliation[a]{School of Fundamental Physics and Mathematical Sciences, Hangzhou Institute for Advanced Study, University of Chinese Academy of Sciences, Hangzhou, Zhejiang, 310024, China}

\affiliation[b]{Beijing Institute of Mathematical Sciences and Applications, Beijing, 101408, China}

\affiliation[c]{Yau Mathematical Sciences Center, Tsinghua University, Beijing, 100084, China}

\affiliation[d]{Institute of Theoretical Physics, Chinese Academy of Sciences, Beijing, 100190, China}

\affiliation[e]{School of Physical Sciences, University of Chinese Academy of Sciences, Beijing, 100049, China}

\emailAdd{chenqian192@mails.ucas.ac.cn}

\emailAdd{lars.andersson@bimsa.cn}

\emailAdd{liliphy@itp.ac.cn}

\abstract{
	Based on the spectral decomposition technique, we introduce a simple and universal numerical method to analyze the stability of solitons.
	Adopting this method, the linear dynamical properties of $Q$-balls are systematically revealed, from the fundamental to the excited states. 
	For the fundamental $Q$-ball, the well-known stability criterion holds.
	However, for the excited $Q$-balls, the situation becomes extremely complicated, in which the stability criterion is violated.
	The system exhibits dynamical instability to both spherically symmetric and non-spherically symmetric perturbations, manifested in the appearance of complex and imaginary modes.
	In addition, we observe two interesting phenomena.
	One is that the oscillation mode and the complex or imaginary mode can transform into each other, marking the transition of the dynamical properties of the system.
	The other is the existence of excited $Q$-balls capable of resisting perturbations with low-order spherical harmonics.
	Such results indicate that the excited $Q$-balls will exhibit rich dynamical behaviors.
}


\arxivnumber{}

\begin{document}
	
\maketitle

%=======================================================================
\section{Introduction}\label{sec:I}
Non-topological solitons \cite{Lee:1991ax,Zhou:2024mea} have attracted significant attention with a wide range of applications in multiple branches of modern physics.
In particle physics, they are found to exist widely in various supersymmetric extensions of the Standard Model \cite{Kusenko:1997zq,Kusenko:1997vi}.
As a candidate for dark matter \cite{Kusenko:1997si,Kusenko:2001vu,Shoemaker:2009kg}, solitons can be copiously produced in the Early Universe.
The study of the formation mechanism and dynamics of solitons has produced a series of consequences in cosmology, such as the problems of
baryon asymmetry \cite{Enqvist:1997si,Enqvist:1998en,Kasuya:1999wu,Kasuya:2000wx} and cosmological phase transitions \cite{Frieman:1988ut,Kusenko:1997hj,Kusenko:2008zm}.
On the other hand, in astrophysics, when the backreaction of matter fields is taken into account, a class of stationary gravitational configurations is derived from solitons, named solitonic boson stars \cite{Lynn:1988rb,Kleihaus:2011sx,Tamaki:2010zz,Collodel:2017biu}, serving as models of compact objects.
Such stars, together with solitons, have important research significance in astronomy, such as mimicking black holes \cite{Troitsky:2015mda,Vincent:2015xta,Vincent:2020dij} and inducing gravitational waves \cite{Palenzuela:2006wp,Palenzuela:2007dm,Kusenko:2008zm,Palenzuela:2017kcg,Cotner:2018vug,Lozanov:2019ylm,Cardoso:2021ehg,Siemonsen:2023hko}.
Beyond asymptotically flat spacetime, within the framework of holographic duality \cite{Maldacena:1997re,Gubser:1998bc,Witten:1998qj,Witten:1998zw}, solitons and boson stars were studied in asymptotically anti-de Sitter spacetime to probe the properties of strongly coupled systems at zero temperature due to the lack of horizon \cite{Astefanesei:2003qy,Nishioka:2009zj,Horowitz:2010jq,Brihaye:2011vk,Hartmann:2012wa,Brihaye:2022oaf,
Rajaraman:2023ygy}.
In addition, motivated by cosmological considerations, the physical properties of these objects in asymptotically de Sitter spacetime were also studied \cite{Palti:2004is,Hartmann:2013kna,Kumar:2016sxx}. 
Interestingly, the interface of solitons was shown to resemble a low-viscosity fluid membrane \cite{Chen:2024axd}, indicating the hydrodynamic properties of solitons.
Hence, the research on solitons and their derivatives is a key node in the development of interdisciplinary physics.
Among them, $Q$-balls \cite{Rosen:1968mfz,Coleman:1985ki}, the spherically symmetric stationary solutions of the self-interacting Klein-Gordon equation, represent the simplest and most representative class.
Investigations into their properties provide valuable insights for understanding other types of solitons.

The physical properties of $Q$-balls in equilibrium have been extensively studied in the literature \cite{Laine:1998rg,Gulamov:2013cra,Shnir:2018yzp,Nugaev:2019vru}.
In the model of a single complex scalar field with a specific attractive self-interaction, $Q$-balls behave as a type of particle-like matter, localized within finite regions of space\cite{Coleman:1985ki}.
Due to the nonlinearity of the stationary equation, the configuration of $Q$-balls can only be obtained analytically for some special scalar potentials \cite{Rosen:1969ay,Bazeia:2015gkq}, such as parabolic ones \cite{Theodorakis:2000bz,Gulamov:2013ema,Nugaev:2013poa,Kovtun:2018jae}.
In the general case, one must resort to some approximate or numerical methods \cite{Coleman:1985ki,Kusenko:1997ad,Multamaki:1999an,PaccettiCorreia:2001wtt,Tsumagari:2008bv,Copeland:2009as,Heeck:2020bau,Heeck:2022iky}.
For a spherically symmetric $Q$-ball, the system can be characterized by a finite energy $E$, a conserved charge $Q$ corresponding to the particle number, and an oscillation frequency $\omega$ interpreted as the chemical potential.
With the additional spatial degree of freedom, spinning $Q$-balls have also been constructed \cite{Volkov:2002aj,Kleihaus:2005me,Kleihaus:2007vk,Brihaye:2008cg,Campanelli:2009su}, whose angular momentum is quantized $J=nQ$ with an integer $n$ and pushes the scalar field away from the center.
In addition, when a gauge field is introduced to generate the global $U(1)$ symmetry into the local $U(1)$ symmetry, the scalar field will carry an electric charge, and the corresponding condensation is known as the gauged $Q$-ball \cite{Lee:1988ag,Anagnostopoulos:2001dh,Arodz:2008nm,Tamaki:2014oha,Gulamov:2015fya,Heeck:2021zvk,Heeck:2021gam}.
The appearance of the gauge field brings additional repulsion to the system, which significantly affects the configuration and stability of $Q$-balls.
In either case, beyond the ground state, there is a series of excited $Q$-balls with additional radial nodes \cite{Friedberg:1976me,Volkov:2002aj,Mai:2012cx,Loginov:2020xoj,Almumin:2021gax}.
Furthermore, considering non-spherical configurations or asymptotic non-vacuum, there are various analogues of $Q$-balls \cite{Axenides:2001pi,Sakai:2010ny,Nugaev:2016wyt}.
The diversity of field configurations enriches this research field and provides more possibilities for physical phenomena.

Beyond the properties of the equilibrium state, the further essential issue is the stability of $Q$-balls \cite{Gleiser:2005iq,Sakai:2007ft,Mai:2012yc,Kovtun:2016qyj,Panin:2016ooo,Smolyakov:2017axd,Smolyakov:2019cld,Loiko:2022noq}, which characterizes the physical properties of the system near equilibrium.
However, because of the nonlinear nature of the field equations, it is a challenge to construct a complete theoretical framework to address such a problem.
At present, the stability of fundamental $Q$-balls has been well understood, which is generally divided into three types in the absence of interactions with fermions \cite{Cohen:1986ct,Clark:2005zc,Kawasaki:2012gk}:
\begin{enumerate}
    \item Quantum mechanical stability \cite{Callan:1977pt,Coleman:1977py,Graham:2001hr,Levkov:2017paj} -- the energy of a $Q$-ball is lower than that of a collection of free particle quanta with the same particle number, that is, $E<mQ$ with the rest mass of the free particle $m$. Consequently, the $Q$-ball is prevented from decaying into free particles.
    \item Classical stability -- the $Q$-ball is stable with respect to small fluctuations, which is proved to be equivalent to the condition $dQ/d\omega<0$ \cite{Friedberg:1976me,Lee:1991ax}. In this situation, all linear-order perturbations will dissipate or oscillate over time on the background of the $Q$-ball, unable to grow.
    \item Stability against fission -- the energy of a single $Q$-ball is lower than the total energy of several $Q$-balls with smaller charges, given the same total charge. This is shown to require $d^{2}E/dQ^{2}<0$ \cite{Lee:1991ax,Gulamov:2013ema}. As a result, the $Q$-ball is prevented from fragmentation.
\end{enumerate}
In addition, due to the basic relation of physical quantities in the relativistic theory of non-topological solitons $dE/dQ=\omega$ \cite{Lee:1991ax,Gulamov:2013cra}, the condition for stability against fission was shown to be identical to that for classical stability.
Therefore, based on the above stability criteria, fundamental $Q$-balls can be classified into three types:
\begin{itemize}
	\item absolutely stable, the strong stability condition (1) is met;
	\item metastable, condition (1) is not met but the weak stability condition (2) is;
	\item unstable, neither conditions (1) and (2) are satisfied.
\end{itemize} 
However, such stability criteria are limited and fail for general $Q$-ball configurations.
For example, for the classical stability of gauged $Q$-balls, due to the complexity of the eigenspectrum of the linear perturbation operator, one can not draw a direct conclusion between the sign of $dQ/d\omega$ and the classical stability \cite{Panin:2016ooo}, as well as for spinning $Q$-balls and excited $Q$-balls.
It is a challenge to analytically obtain the classical stability criteria for these general $Q$-balls, even in the simple case of spherically symmetric perturbations.
In this case, an alternative approach is to dynamically evolve the perturbed configurations by numerically solving the nonlinear field equations \cite{Kinach:2022jdx,Kinach:2024hfa,Kinach:2024qzc}.
All in all, the construction of a theoretical framework for soliton stability still requires more efforts and attempts, either analytically or with the help of numerical methods.
Such a framework would be of great significance for the non-equilibrium dynamics and physical applications of solitons \cite{Axenides:1999hs,Battye:2000qj,Amin:2019ums,Hong:2024uxl}.

In this paper, using numerical methods to directly solve the linearized self-interacting Klein-Gordon equation, we systematically study the dynamical stability\footnote{The dynamical stability here is essentially the classical stability mentioned earlier, characterized by the spectrum of the linearized operator of the field equations. If there exists an unstable mode that grows exponentially with time, the system is said to be dynamically unstable at the linear level, or to exhibit mode instability. Otherwise, the system is stable.} of $Q$-balls from the fundamental to the excited states.
On the one hand, intuitively, the excited state is usually dynamically unstable and spontaneously undergoes a dynamical transition to the ground state.
However, there are some studies showing that excited boson stars can be dynamically stable within a specific parameter range, depending on the self-interaction of matter \cite{Sanchis-Gual:2021phr,Brito:2023fwr,Nambo:2024gvs}.
A natural question is whether there are dynamically stable excited $Q$-balls.
This work uncovers the existence of such excited states, suggesting that they possess rich dynamical behaviors.
On the other hand, numerical methods based on field discretization can directly solve the eigenvalue problem generated by linear perturbation analysis.
Such a simple and efficient strategy is applicable to almost all types of perturbation equations.
Therefore, it is expected to become a universal method for analyzing the dynamical stability of $Q$-balls and other general solitons.
We hope that this research will be helpful in constructing the theoretical framework for the stability of solitons. 

The organization of the paper is as follows.
In section \ref{sec:Q}, we introduce the solitonic field theory with a global $U(1)$ symmetry and show the equilibrium properties of $Q$-balls and their radial excitations.
In section \ref{sec:Lpa}, we present the numerical strategy for analyzing the dynamical stability of solitons, including perturbation equations, boundary conditions, numerical setup, and numerical convergence.
In section \ref{sec:Ds}, we systematically analyze the linear dynamical properties of $Q$-balls and their radial excitations.
For the fundamental $Q$-ball, the situation is simple, where the stability criterion is applicable. 
However, for the excited $Q$-balls, the stability criterion no longer applies and the situation is extremely complicated, with the emergence of imaginary and complex modes characterizing the dynamical instability and configurations capable of resisting perturbations with low-order spherical harmonics.
Such results indicate that the excited $Q$-balls will exhibit rich dynamical behaviors, such as dynamical transitions between them and the fundamental $Q$-ball, free particles. We end this paper with a summary and an outlook in section \ref{sec:C}.
%=======================================================================

%=======================================================================
\section{Global $U(1)$ theory}\label{sec:Q}
In this section, we introduce the global $U(1)$ theory from the setup of the theoretical model, the field equations, and the spherically symmetric stationary solutions -- $Q$-balls and their radial excitations.

\subsection{Model and setup}
The solitonic field theory with a global $U(1)$ symmetry involves a self-interacting complex scalar field, described by the following Lagrangian density in (d+1) dimensional spacetime.
\begin{equation}
	\mathcal{L}=-\nabla_{\mu}\psi\nabla^{\mu}\psi^{*}-V(|\psi|),\label{eq:2.1}
\end{equation}
where $\nabla$ is the covariant derivative associated with the spacetime metric $g_{\mu\nu}$, $V(|\psi|)$ is a $U(1)$ invariant scalar field potential.
The self-interacting Klein-Gordon equation governing the system is given by
\begin{equation}
	\nabla_{\mu}\nabla^{\mu}\psi=\frac{\partial V(|\psi|)}{\partial\psi^{*}},\label{eq:2.2}
\end{equation}
together with the energy-momentum tensor 
\begin{equation}
	T_{\mu\nu}=\nabla_{\mu}\psi\left(\nabla_{\nu}\psi\right)^{*}+\nabla_{\nu}\psi\left(\nabla_{\mu}\psi\right)^{*}+g_{\mu\nu}\mathcal{L}\,.
\end{equation}
One has the Noether current associated with the $U(1)$ symmetry of the theory
\begin{equation}
	J_{\mu}=i\left[\psi^{*}\nabla_{\mu}\psi - \psi\left(\nabla_{\mu}\psi\right)^{*}\right].
\end{equation}
By integrating the energy density and charge density over the entire space, one can further obtain the conserved energy and conserved charge of the system respectively
\begin{equation}
	E=\int_{\Sigma}  T_{\mu\nu}\xi^{\mu}n^{\nu}\sqrt{\gamma}d^{d}x,\quad  Q=\int_{\Sigma}  J_{\mu}n^{\mu}\sqrt{\gamma}d^{d}x,
\end{equation}
where $\Sigma$ represents a $d$-dimensional space-like asymptotically flat hypersurface bounded at spatial infinity, with the induced metric $\gamma_{\mu\nu}$, the time-like Killing vector $\xi^{\mu}$ and the future-directed unit normal vector $n^{\mu}$.

To ensure that $\psi=0$ is a true vacuum and the mass of the matter field is positive, the scalar potential should satisfy
\begin{equation}
	V(\psi=0)=0,\quad V(\psi\neq0)>0,\quad \frac{d^{2} V}{d|\psi|^{2}}(\psi=0)>0.
\end{equation}
Furthermore, in order to allow for the existence of non-trivial solutions -- $Q$-balls, it has been shown that the growth rate of the scalar potential should be slower than that of the quadratic mass term \cite{Coleman:1985ki}.
Such a requirement can be achieved in the polynomial form by adding a higher-order damping term.
Among them, we focus on the simplest type of polynomial potential with equidistant exponents
\begin{equation}
	V=m^{2}|\psi|^{2}-\lambda^{2}|\psi|^{3}+\kappa|\psi|^{4},\label{eq:2.7}
\end{equation}
where $m$ is the mass of the scalar field and the highest order coefficient satisfies $\kappa>\frac{\lambda^{4}}{4m^{2}}$ so that the potential has a unique zero, namely the vacuum $\psi=0$.
On the other hand, one can further remove two of the parameters by rescaling the units of coordinates and energy.
Using the dimensionless variables
\begin{equation}
	m x\rightarrow x,\quad \lambda^{2}\psi/m^{2}\rightarrow \psi,\quad m^{2}\kappa/\lambda^{4}\rightarrow \kappa,
\end{equation}
the scalar potential (\ref{eq:2.7}) is reduced to 
\begin{equation}
	V=|\psi|^{2}-|\psi|^{3}+\kappa|\psi|^{4},\label{eq:2.9}
\end{equation}
with $\kappa> 1/4$.
In what follows, we will fix the above single parameter to be $\kappa=1$ for definiteness.
In order to rule out peculiarities, we have analyzed multiple cases with coupling strengths within range $0.28\leq\kappa\leq 1.25$ and reached the same conclusion at a qualitative level.

\subsection{$Q$-balls and radial excitations}
In this paper, we work in ($3+1$)-dimensional Minkowski spacetime with the line element
\begin{equation}
	ds^{2}=-dt^{2}+dr^{2}+r^{2}\left(d\theta^{2}+\sin(\theta)^{2}d\varphi^{2}\right),
\end{equation}
and focus on the spherically symmetric $Q$-balls with the ansatz
\begin{equation}
	\psi(t,r)=e^{-i\omega t}\phi(r),
\end{equation}
where the oscillation frequency $\omega$ is a real positive constant and $\phi(r)$ is a real radial profile function determined by the following differential equation induced from (\ref{eq:2.2}):
\begin{equation}
	\frac{d^{2}\phi}{dr^{2}} + \frac{2  }{r}\frac{d\phi}{dr} + \left(\omega^{2}  - V'\right) \phi =0,\label{eq:2.12}
\end{equation}
with $V'= \frac{dV}{d|\psi|^{2}}(\phi)$.
By taking an asymptotic expansion of the above equation near the origin and infinity in space, one can obtain the asymptotic behaviors of $\phi$ as follows
\begin{subnumcases}{\phi(r)=}
	\phi_{0}+\frac{\phi_{0}}{6}\left(V'(\phi_{0})-\omega^{2}\right)r^{2}+\cdots,&$r\rightarrow 0$,\label{eq:2.13a}\\
	\frac{\phi_{\infty}}{r}\exp{\left(-r\sqrt{m^{2}-\omega^{2}}\right)}+\cdots,
	&$r\rightarrow \infty$,\label{eq:2.13b}
\end{subnumcases}
where the constants $\phi_{0}$ and $\phi_{\infty}$ can only be determined after solving the radial equation (\ref{eq:2.12}).
From the above asymptotic behavior (\ref{eq:2.13b}), the finiteness of the energy of the system implies
\begin{equation}
	\omega^{2}<m^{2},\label{eq:2.14}
\end{equation}
so that the amplitude of the scalar field decays exponentially at large $r$, ensuring that the solution is a bound state.

In order to search for $Q$-ball solutions by shooting method, the radial equation (\ref{eq:2.12}) is interpreted as the Newtonian equation describing the motion of a classical particle of unit mass, with position $\phi$ and time $r$, in the effective potential 
\begin{equation}
	U=\frac{1}{2}\left(\omega^{2}\phi^{2}-V\right),
\end{equation}
and under the influence of the friction proportional to the ratio of velocity to time $F=-\frac{2}{r}\frac{d\phi}{dr}$.
In this consideration, a $Q$-ball corresponds to a trajectory that starts from position $\phi=\phi_{0}$ at time $r=0$ and terminates at origin $\phi=0$ after infinite time $r\rightarrow\infty$.
Due to the existence of friction, the mechanical energy $E_{m}=\frac{1}{2}\left(\frac{d\phi}{dr}\right)^{2}+U$ of the particle continues to decrease during the motion.
Therefore, the possibility of reaching the end point depends on the particle possessing more mechanical energy at the starting point than at the end point.
Such a requirement gives a lower bound on the oscillation frequency
\begin{equation}
	\omega^{2}>\min\left[\frac{V}{\phi^{2}}\right]=m^{2}-\frac{\lambda^{4}}{4\kappa}.\label{eq:2.16}
\end{equation}
Combining conditions (\ref{eq:2.14}) and (\ref{eq:2.16}), for the rescaled self-interaction potential (\ref{eq:2.9}) considered in this paper, the existence condition for $Q$-ball solutions is
\begin{equation}
	1-\frac{1}{4\kappa}=\omega^{2}_{-}<\omega^{2}<\omega^{2}_{+}=1,\label{eq:2.17}
\end{equation}
where the upper and lower bounds correspond to the thick-wall and thin-wall limits, respectively.

%%%%%%%%%%%%%%%%%%
\begin{figure}
	\centering
	\includegraphics[width=.49\linewidth]{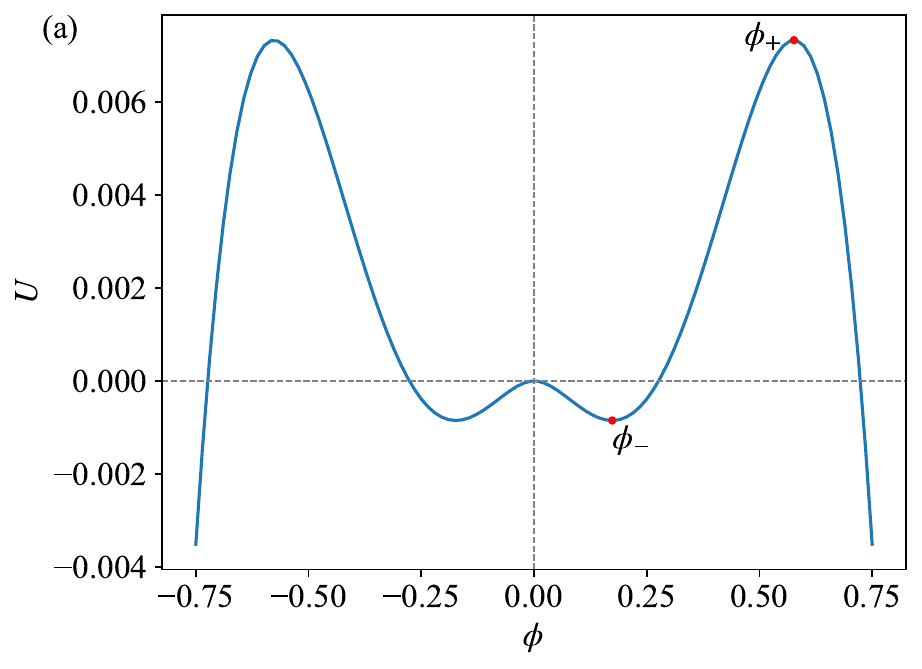}
	\includegraphics[width=.49\linewidth]{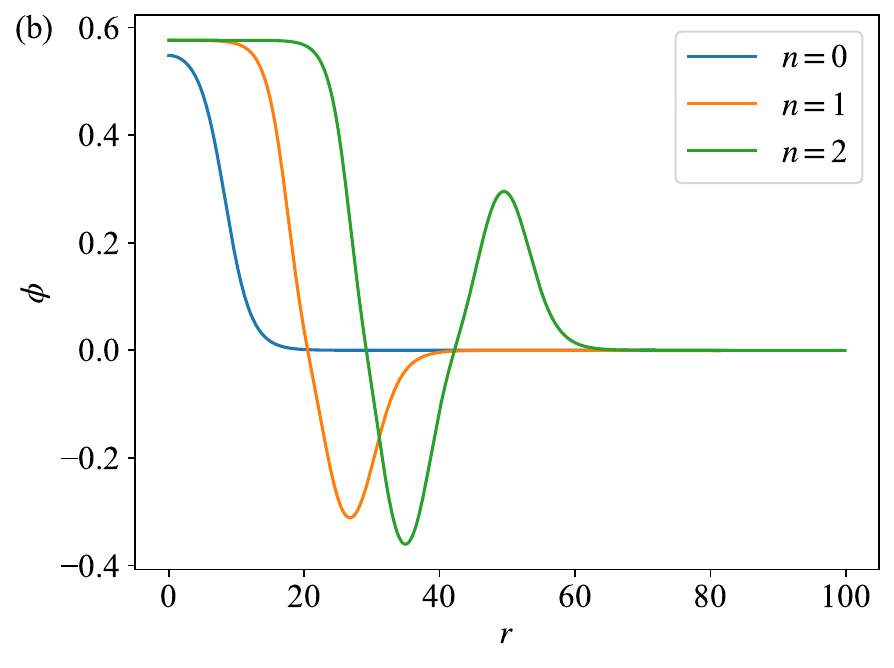}
	\caption{(a) The effective potential $U$ with $\omega^{2}=0.8$. The extrema $\phi_{\pm}$ are shown as red dots. (b) The configurations of the fundamental $Q$-ball ($n=0$) and the first two excited states ($n=1,2$).}
	\label{fig:1-2}
\end{figure}
%%%%%%%%%%%%%%%%%%

The left panel of Figure \ref{fig:1-2} gives an example of the effective potential $U$ with $\omega^{2}=0.8$.
For a particle with an initial position at $\phi_{0}\in\left(\phi_{-},\phi_{+}\right)$, there are three possibilities for the final position\footnote{From the asymptotic behavior (\ref{eq:2.13a}), it can be seen that the initial velocity of the particle is zero $\frac{d\phi}{dr}(r=0)=0$, thus it can not escape the energy well by crossing the maxima $\pm\phi_{+}$ in the effective potential.
On the other hand, due to the existence of friction, the particle will inevitably settle down to a local extreme of the effective potential.}: the origin $\phi=0$ and the extrema $\pm\phi_{-}$, among which only the trajectory with the origin as the end point corresponds to the $Q$-ball solution.
As the initial position $\phi_{0}$ gradually increases from the minimum $\phi_{-}$, a critical value will be encountered, below which the particle converges to the minimum $\phi_{-}$ with oscillation, otherwise (slightly larger than the critical value) it crosses the origin and is attracted to the other minimum $-\phi_{-}$.
The trajectory with this critical value as the initial position is exactly the fundamental $Q$-ball, as shown by the blue curve in the right panel of Figure \ref{fig:1-2}.
Continuing to increase the initial position $\phi_{0}$ to obtain more mechanical energy, the second critical value corresponding to the first excited state will appear.
In this case, after crossing the origin, the particle rushes uphill until it reaches a point of return, then approaches the origin from the negative region and stays there.
Such a trajectory indicates that the profile function of the first excited $Q$-ball possesses a radial node.
Similarly, the closer the initial position $\phi_{0}$ is to the maximum $\phi_{+}$, the more times the particle overshoots the origin.
The resulting trajectories correspond to the excited states with more radial nodes $n$.
In this paper we focus on the fundamental $Q$-ball ($n=0$) and the first two excited states ($n=1,2$).

%%%%%%%%%%%%%%%%%%
\begin{figure}
	\centering
	\includegraphics[width=.49\linewidth]{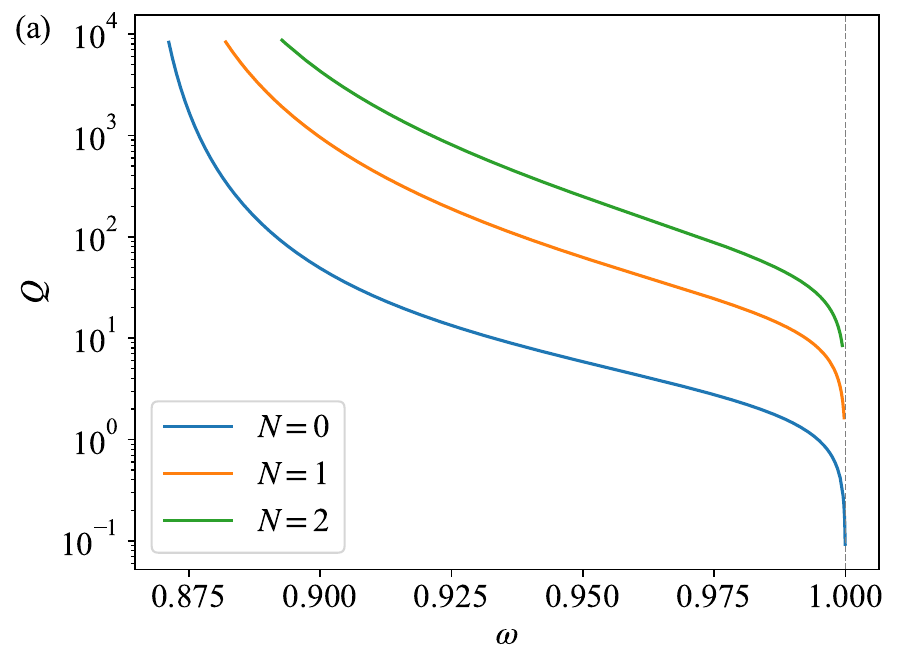}
	\includegraphics[width=.49\linewidth]{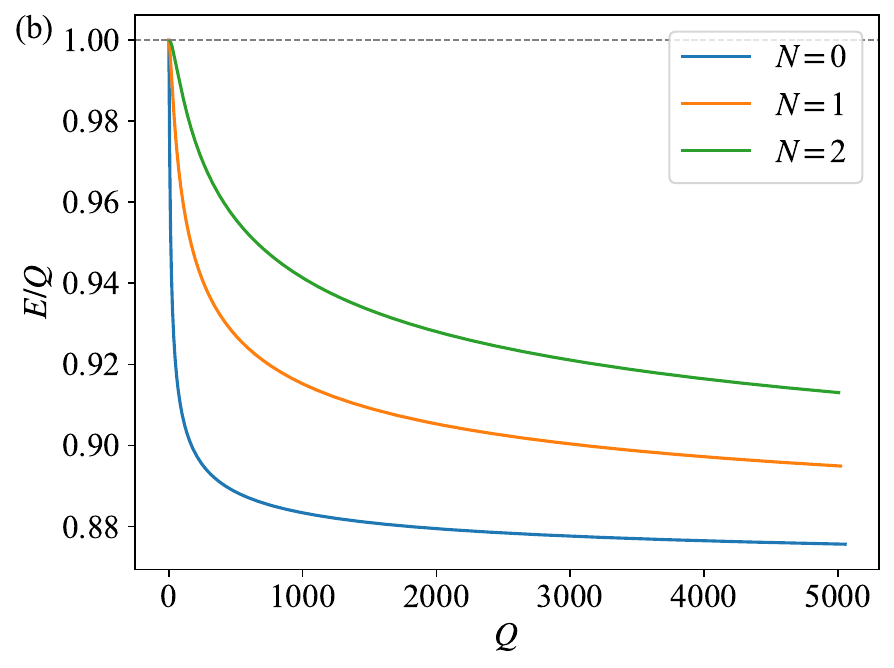}
	\caption{The relationship between the physical quantities of $Q$-balls. (a) The charge $Q$ as a function of oscillation frequency $\omega$. (b) The ratio of energy to charge $E/Q$ as a function of charge $Q$. Curves of different colors represent fundamental ($n=0$) and excited ($n=1,2$) $Q$-balls respectively. }
	\label{fig:3-4}
\end{figure}
%%%%%%%%%%%%%%%%%%

Figure \ref{fig:3-4} shows the relationship between the physical quantities of $Q$-balls.
The energy and charge approach a minimum value of zero at the thick-wall limit, with the ratio converging to the value of the oscillation frequency at the boundary $\omega_{+}$.
The reason for such a ratio is that the relation $dE/dQ=\omega$ holds \cite{Lee:1991ax,Gulamov:2013cra}.
At the thin-wall limit, these physical quantities grow infinitely, and it can be expected that their ratio should be equal to the value of the oscillation frequency at the other boundary $\omega_{-}$, which can be verified analytically but is difficult for numerical methods \cite{Heeck:2020bau}.
In addition, the monotonically decreasing charge (left panel of Figure \ref{fig:3-4}) indicates that the classical stability criterion $\frac{dQ}{d\omega}<0$ is satisfied.
On the other hand, the fact that the ratio of energy to charge is less than the mass of the scalar field indicates that the quantum stability criterion $E<mQ$ is also satisfied.
Therefore, from the perspective of the stability criterion, all $Q$-balls are absolutely stable.
However, for excited $Q$-balls, such stability analysis should not be applicable, although the structural characteristics of the phase diagram are similar to those of the fundamental $Q$-ball.
Since the energy of excited states is always higher than that of the fundamental state in the case of fixed charge, and increases with the number of radial nodes $n$, all excited $Q$-balls are expected to be dynamically unstable and can transition to the fundamental $Q$-ball.
%=======================================================================

%=======================================================================
\section{Linear perturbation analysis}\label{sec:Lpa}
In this section, we apply a perturbation to the scalar field on the background of the stationary configurations obtained above $\psi=e^{-i\omega t}\phi +\delta\psi$ to perform a linear stability analysis, including perturbation equations, boundary conditions, numerical method, and numerical convergence.

\subsection{Perturbation equations}
Within the theoretical framework of linear perturbation analysis, what we need to deal with is the following linearized Klein-Gordon equation that dominates the properties of the perturbation:
\begin{equation}
	\left[-\partial^{2}_{t}+\Delta-V'(\phi)\right]\delta\psi-\phi^{2}V''(\phi)\left(\delta\psi+e^{-2i\omega t}\delta\psi^{*}\right)=0, \label{eq:3.1}
\end{equation}
where the prime represents the derivative with respect to the variable $|\psi|^{2}$, and the symbol $\Delta$ is the three-dimensional Laplace operator.
Due to the existence of self-interaction, the perturbation function $\delta\psi$ is coupled with its conjugate $\delta\psi^{*}$, resulting in the monochromatic wave failing to solve the above perturbation equation.
The correct ansatz for the perturbation should contain at least one pair of dichromatic waves. \footnote{Such an ansatz does not cover all possible forms of perturbations, such as the modes corresponding to the Lorentz transformations \cite{Smolyakov:2017axd}.}
\begin{equation}
	\delta\psi=\sum_{l,m}\left(e^{-i\left(\omega +\Omega\right) t}\delta \psi^{(l,m)}_{+}\left(r\right) +e^{-i\left(\omega - \Omega^{*}\right) t} \delta \psi^{(l,m)}_{-}\left(r\right)\right)Y_{(l,m)}\left(\theta,\varphi\right), \label{eq:3.2}
\end{equation}
where the radial perturbation functions $\delta\psi^{(l,m)}_{\pm}$ are the coefficients of the eigenstates associated with the eigenfrequency $\Omega=\Omega_{R}+i\Omega_{I}$ in the expansion based on spherical harmonics $Y_{(l,m)}$.
For a mode with a positive imaginary part of the eigenfrequency $\Omega_{I}>0$, the mode will grow exponentially over time in the form of $e^{\Omega_{I}t}$, thereby pushing the system away from the equilibrium state.
Such a mode is dynamically unstable.
In contrast, a mode with a negative imaginary part of the eigenfrequency $\Omega_{I}<0$ will decay exponentially, failing to trigger instability in the system.

Substituting the perturbation ansatz (\ref{eq:3.2}) into the linearized Klein-Gordon equation (\ref{eq:3.1}) and defining new radial perturbation functions $\delta\psi^{(l)}_{\pm}=\sum_{m=-l}^{l}\delta\psi^{(l,m)}_{\pm}$, one can obtain the following perturbation equations 
\begin{equation}
	\left[\Omega^{2}\begin{pmatrix}
		1&0\\
		0&1
	\end{pmatrix}
	+2\omega\Omega\begin{pmatrix}
		1&0\\
		0&-1
	\end{pmatrix}
	+\begin{pmatrix}
		\mathbb{L}^{(l)}&-\phi^{2}V''(\phi)\\
		-\phi^{2}V''(\phi)&\mathbb{L}^{(l)}
	\end{pmatrix}\right]
	\begin{pmatrix}
		\delta\psi^{(l)}_{+}\\
		\delta\psi^{(l)*}_{-}
	\end{pmatrix}=0,\label{eq:3.3}
\end{equation}
with the differential operator
\begin{equation}
	\mathbb{L}^{(l)}=\frac{d^{2}}{dr^{2}}+\frac{2}{r}\frac{d}{dr}-\frac{l\left(l+1\right)}{r^{2}}+\omega^{2}-V'(\phi)-\phi^{2}V''(\phi),
\end{equation}
from which it can be clearly observed that the two waves are coupled.
Further, by taking an asymptotic expansion of the above perturbation equations near the origin and infinity, one can obtain the asymptotic behaviors of the radial perturbation functions as follows
\begin{subnumcases}{\delta\psi^{(l)}_{\pm}=\label{eq:3.5}}
		A_{\pm}r^{l}+\cdots,&$r\rightarrow$ 0,\label{eq:3.5a}\\
		\frac{1}{r}\left(B_{\pm}e^{-k_{\pm}r}+ C_{\pm}e^{k_{\pm}r}\right)+\cdots,
		&$r\rightarrow \infty$,\label{eq:3.5b}
\end{subnumcases}
with constants $\{A_{\pm},B_{\pm},C_{\pm}\}$ and exponential coefficients
\begin{equation}
	k_{+}=\sqrt{m^{2}-\left(\omega+\Omega\right)^{2}},\quad k_{-}=\sqrt{m^{2}-\left(\omega-\Omega^{*}\right)^{2}}.
\end{equation}
Without loss of generality, we conventionally take the real part of the coefficients $k_{\pm}$ to be positive.
Under such a convention, the branches with coefficients $B_{\pm}$ in the asymptotic behavior (\ref{eq:3.5b}) converge exponentially, showing a bound state, while the other branches with coefficients $C_{\pm}$ diverge.
On the other hand, combined with (\ref{eq:3.2}), one finds that Im$[k_{\pm}]\sim-\left(\omega\pm\Omega_{R}\right)\Omega_{I}$, thus the propagation direction of the perturbation waves at infinity is uniquely determined by the sign of the imaginary part of the eigenfrequency $\Omega_{I}$, or equivalently, the stability of the mode.
For a stable mode with $\Omega_{I}<0$, the branches with coefficients $B_{\pm}$ represent the incident waves, and the other branches with coefficients $C_{\pm}$ represent the outgoing waves.
The situation is exactly the opposite for an unstable mode with $\Omega_{I}>0$, where the branches with coefficients $B_{\pm}$ and $C_{\pm}$ are the outgoing and incident waves, respectively.
The general case of the complex eigenfrequency described above is summarized in Table \ref{table:1}.
For the special case of the real eigenfrequency, according to its value, the asymptotic behavior (\ref{eq:3.5b}) can be distinguished into three types, as shown in Table \ref{table:2}.
Firstly, for the case of low eigenfrequency $0<\Omega<m-\omega$, since the exponential coefficients $k_{\pm}$ are real numbers, the branches with coefficients $B_{\pm}$ and $C_{\pm}$ converge and diverge respectively, similar to the above situation, except that there is no wave component.
Secondly, for the case of intermediate eigenfrequency $m-\omega<\Omega<m+\omega$, the perturbation $\delta\psi_{+}$ takes the form of pure waves with the incident amplitude $B_{+}$ and the outgoing amplitude $C_{+}$, and the perturbation $\delta\psi_{-}$ is still the superposition of the bound state $B_{-}$ and the divergent solution $C_{-}$.
Finally, for the case of high eigenfrequency $m+\omega<\Omega$, the exponential coefficients $k_{\pm}$ are both pure imaginary numbers, so the perturbations $\delta\psi_{\pm}$ both exist in the form of waves, with the incident and outgoing amplitudes being $\{B_{+},C_{-}\}$ and $\{C_{+},B_{-}\}$ respectively. 
Based on the above situations, there are several types of possible boundary conditions, of which the following three are widely used according to different physical motivations:
\begin{itemize}
	\item Scattering boundary condition -- If the eigenfrequency is real and in the high frequency region $\Omega>m+\omega$, as mentioned above, such a physical scenario describes a type of scattering problem of dichromatic waves with frequencies $\omega\pm\Omega$ on the background of a $Q$-ball \cite{Saffin:2022tub,Cardoso:2023dtm,Zhang:2024ufh}, where the amplitudes of the incident and outgoing waves are $\{B_{+},C_{-}\}$ and $\{C_{+},B_{-}\}$ respectively.
	Such research can allow us to reveal the energy transfer and superradiance phenomena of the system.
	However, since the scattering frequency $\Omega$ is restricted to be real, it can not characterize the dynamical stability of the system.
	\item Quasi-normal mode (QNM) boundary condition -- This case, defined in a purely dissipative system, requires that the perturbation contains only the branch of the outgoing waves at infinity \footnote{For a black hole system, in addition to the purely outgoing-wave boundary condition at infinity, the QNM problem also requires that there are only ingoing waves at the event horizon to satisfy the requirement that classically nothing should leave the interior of the black hole.}, without incoming radiations \cite{Berti:2009kk,Konoplya:2011qq}.
	\item Bound state (BS) boundary condition -- It is defined in the perturbation problem of a massive matter field \cite{Lasenby:2002mc,Dolan:2007mj,Rosa:2011my,Pani:2012bp,Macedo:2013jja}, and requires that the perturbation eigenstates are spatially localized near the perturbed object and decay exponentially at infinity, which in our case is
	\begin{equation}
		C_{\pm}=0.\label{eq:3.7}
	\end{equation}
	
\end{itemize}

The specific implementations of the last two boundary conditions are summarized in Tables \ref{table:1} and \ref{table:2}.
For both boundary conditions, the eigenfrequency $\Omega$ usually exhibits an infinite discrete distribution on the complex plane, which only depends on the parameters of the stationary background, characterizing the intrinsic dynamical stability of the system.
Therefore, this is exactly the issue that needs to be considered in this work.
    
\begin{table}[h!]
	\centering
	\begin{tabular}{c|c|c|c|c|c}
		\hline
		&$\Omega_{I}$&waves $B_{\pm}$&waves $C_{\pm}$&QNM&BS\\
		\hline
		\hline
		Stable&$\Omega_{I}<0$&convergent, ingoing&divergent, outgoing&$B_{\pm}=0$&$C_{\pm}=0$\\
		\hline
		Unstable&$\Omega_{I}>0$&convergent, outgoing&divergent, ingoing&$C_{\pm}=0$&$C_{\pm}=0$\\
		\hline
	\end{tabular}
	\caption{QNM boundary condition and BS boundary condition for the perturbation with a complex eigenfrequency $\Omega=\Omega_{R}+i\Omega_{I}$ of a massive scalar field on the background of a $Q$-ball.}
	\label{table:1}
\end{table}

\begin{table}[h!]
	\centering
	\begin{tabular}{c|c|c|c|c|c|c}
		\hline
		$\Omega_{R}$&wave $B_{+}$&wave $C_{+}$&wave $B_{-}$&wave $C_{-}$&QNM&BS\\
		\hline
		\hline
		$\left(0,m-\omega\right)$&convergent&divergent&convergent&divergent&$C_{\pm}=0$&$C_{\pm}=0$\\
		\hline
		$\left(m-\omega,m+\omega\right)$&ingoing&outgoing&convergent&divergent&$B_{+},C_{-}=0$&-\\
		\hline
		$\left(m+\omega,\infty\right)$&ingoing&outgoing&outgoing&ingoing&$B_{+},C_{-}=0$&-\\
		\hline
	\end{tabular}
	\caption{QNM boundary condition and BS boundary condition for the perturbation with a real eigenfrequency $\Omega=\Omega_{R}$ of a massive scalar field on the background of a $Q$-ball.}
	\label{table:2}
\end{table}
	
\subsection{Spectral decomposition}
For linear perturbation problems, many solution strategies have been developed, such as the WKB approximation \cite{Schutz:1985km,Iyer:1986np,Konoplya:2003ii} suitable for the perturbation equation with Schr$\ddot{\text{o}}$dinger-like form and the continued fraction method \cite{Leaver:1985ax,Leaver:1986vnb,Leaver:1986gd,Leaver:1990zz,Nollert:1993zz} based on the Frobenius series.
The applicability of these methods depends on the characteristics of the perturbation equation and the stationary background.
For our situation here, the entanglement between the dichromatic waves $\delta\psi_{\pm}$ makes the perturbation equation irreducible to the Schr$\ddot{\text{o}}$dinger-like form.
On the other hand, since the background configuration $\phi(r)$ can only be obtained numerically, the recursive relationship between the coefficients of the Frobenius series is difficult to express.
Such features make traditional methods difficult to implement.
An alternative is to adopt a numerical discretization approach based on spectral decomposition \cite{Chung:2023zdq,Chung:2023wkd}.
The applicability of this method is almost unrestricted by the form of the perturbation equations and by whether the background configuration is analytical or numerical, so it is expected to be developed into a general method for analyzing linear perturbation problems.

The only difficulty in adopting the spectral decomposition method is the imposition of boundary conditions, especially the QNM boundary condition.
For a massless matter field that propagates with unit velocity at infinity, one can reduce the perturbation equation, which degenerates into a free wave equation near infinity, into the advection equation and then filter out the incident waves by setting the sign of the unit velocity, or equivalently adopting a spacetime slicing that intersects future null infinity \cite{Zenginoglu:2011jz,Ansorg:2016ztf}.
However, for a massive matter field, since its propagation velocity at infinity depends on the eigenfrequency, such a strategy fails.
Moreover, the coupling of dichromatic waves makes the boundary condition more complicated, as shown in Tables \ref{table:1} and \ref{table:2}.
Fortunately, if one only considers the dynamical stability of the system, that is, looking for unstable modes with $\Omega_{I}>0$, then the QNM boundary condition coincides with the BS boundary condition.
Furthermore, since the other branch coexisting with the bound state must be divergent, the BS boundary condition (\ref{eq:3.7}) can be simply equivalent to requiring the perturbation functions $\delta\psi_{\pm}$ to be regular at infinity.
Therefore, in this work, we adopt the spectral decomposition method to numerically solve the perturbation equations (\ref{eq:3.3}) with the regular boundary condition (\ref{eq:3.7}).
In this case, for each mode $\{\Omega,\delta\psi=\left(\delta\psi_{+},\delta\psi^{*}_{-}\right)\}$, there are three other coexisting solutions $\{\Omega^{*},\delta\psi=\left(\delta\psi^{*}_{+},\delta\psi_{-}\right)\}$, $\{-\Omega,\delta\psi=\left(\delta\psi^{*}_{-},\delta\psi_{+}\right)\}$, and $\{-\Omega^{*},\delta\psi=\left(\delta\psi_{-},\delta\psi^{*}_{+}\right)\}$, indicating that the eigenfrequency $\Omega$ presents a highly symmetric distribution on the complex plane.
On the other hand, since there are no bound states for the real eigenfrequency within the region $\Omega>m-\omega$, we only focus on the modes located in the first quadrant, the upper part of the imaginary axis, and the low-frequency region on the real axis.
Such modes are also QNMs.

For simplicity, we adopt the equivalent form of spectral decomposition, the pseudo-spectral discretization, which is more suitable for the calculation of algebraic systems.
Overall, such an approach can be broken down into three steps.
The first step is the preprocessing of the perturbation equations, including the compactification of coordinates and the redefinition of field functions.
Since infinity is undesirable for numerical methods, we need to restrict the computational domain to a finite range through a coordinate transformation
\begin{equation}
	z=\frac{r}{r+1}\in\left[0,1\right].
\end{equation}
On the other hand, in order for the functions to be solved to adaptively meet the boundary conditions (\ref{eq:3.5}) and(\ref{eq:3.7}), we redefine the perturbation functions as follows
\begin{equation}
	\delta\psi^{(l)}_{\pm}=z^{l}\left(1-z\right)\delta\widetilde{\psi}^{(l)}_{\pm},
\end{equation}
and require the new objective functions $\delta\widetilde{\psi}^{(l)}_{\pm}$ to be regular at the origin and infinity.
Such an approach converts the specific boundary conditions into regular boundary conditions.
The second step is the selection of the spectral collocation grid, which depends on the type of boundary conditions.
For non-periodic boundary conditions, the Chebyshev spectrum is an effective discretization method, which is equivalent to expanding the field functions based on the Chebyshev polynomials \cite{boyd2001chebyshev}.
For specific implementation, there are two common types of grids:
the Chebyshev–Gauss grid
\begin{equation}
	z_{i}=\frac{1}{2}\left[1+\cos\left(\frac{\left(2i-1\right)\pi}{2N}\right)\right],\quad i=1,\cdots,N,
\end{equation}
and the Chebyshev–Gauss–Lobatto grid
\begin{equation}
	z_{i}=\frac{1}{2}\left[1+\cos\left(\frac{i\pi}{N-1}\right)\right],\quad i=0,\cdots,N-1.
\end{equation}
Since the latter contains the interval boundaries $z=\pm 1$, which facilitates the imposition of boundary conditions, it is often used to deal with boundary value problems for partial differential equations.
However, for the regular boundary condition, since the polynomial expansion already requires that the function is regular everywhere in the computational domain, there is no need to impose additional boundary conditions.
In this case, the former is also applicable and more convenient.
This is because for the latter, due to the existence of the $z^{-1}$ term, the discrete equation at the origin must be replaced with the corresponding boundary condition to ensure the regularity of the equations.
In this work, both of the above collocation grids have been tried, and the results agree with each other within the numerical accuracy.
The last step is to numerically solve the resulting eigenvalue problem.
After processing through the above steps, the problem faced is converted into solving a set of algebraic equations with $2N+1$ unknown numbers $\left\lbrace \Omega,\delta\widetilde{\psi}^{(l)}=\left(\delta\widetilde{\psi}^{(l)}_{+}(z_{i}),\delta\widetilde{\psi}^{(l)*}_{-}(z_{i})\right)\right\rbrace$
\begin{equation}
	\left[\Omega^{2}\begin{pmatrix}
		\mathbf{1}&\mathbf{0}\\
		\mathbf{0}&\mathbf{1}
	\end{pmatrix}+2\omega\Omega\begin{pmatrix}
		\mathbf{1}&\mathbf{0}\\
		\mathbf{0}&-\mathbf{1}
	\end{pmatrix} +\mathbf{M}^{(l)}\right]_{2N\times 2N}\delta\widetilde{\psi}^{(l)}=\mathbf{0}_{2N},
\end{equation}
where the matrix $\mathbf{M}$ contains the compactization coordinate $z_{i}$, the differentiation matrix $D_{ij}$ corresponding to the spectral grid, and the background function $\phi(z_{i})$, which dominates the properties of the perturbation equations.
Furthermore, defining $\delta\Phi=\left(\delta\widetilde{\psi},\Omega\delta\widetilde{\psi}\right)$ to reduce the order of the eigenfrequency $\Omega$, the above algebraic equations eventually degenerate into an eigenvalue problem
\begin{equation}
	\begin{pmatrix}
		\mathbf{0}&\mathbf{1}\\
		-\mathbf{M}^{(l)}&2\omega\begin{pmatrix}
			-\mathbf{1}&\mathbf{0}\\
			\mathbf{0}&\mathbf{1}
		\end{pmatrix}_{2N\times 2N}
	\end{pmatrix}_{4N\times 4N}\delta\Phi^{(l)}=\Omega\delta\Phi^{(l)},
\end{equation}
which can be easily solved with a code library such as scipy.

\subsection{Numerical convergence}
In order to ensure the reliability of the results, we need to verify the convergence of the numerical method.
Before that, let us review the restrictions on the form of perturbations.
Using the notations
\begin{equation}
	\begin{pmatrix}
		\delta\Psi^{(l)}_{+}\\
		\delta\Psi^{(l)}_{-}
	\end{pmatrix}=
	\begin{pmatrix}
		1&1\\
		1&-1
	\end{pmatrix}
	\begin{pmatrix}
		\delta\psi^{(l)}_{+}\\
		\delta\psi^{(l)*}_{-}
	\end{pmatrix},
\end{equation}
the perturbation equations (\ref{eq:3.3}) are reduced to the following form
\begin{equation}
	\left[\Omega^{2}\begin{pmatrix}
		1&0\\
		0&1
	\end{pmatrix}
	+2\omega\Omega\begin{pmatrix}
		0&1\\
		1&0
	\end{pmatrix}
	+\begin{pmatrix}
		\mathbb{L}^{(l)}_{+}&0\\
		0&\mathbb{L}^{(l)}_{-}
	\end{pmatrix}\right]
	\begin{pmatrix}
		\delta\Psi^{(l)}_{+}\\
		\delta\Psi^{(l)}_{-}
	\end{pmatrix}=0,\label{eq:3.15}
\end{equation}
with the operators $\mathbb{L}^{(l)}_{\pm}=\mathbb{L}^{(l)}\mp \phi^{2}V''(\phi)$.
As demonstrated in the Appendix A of \cite{Panin:2016ooo}, the mode must satisfy one of the following two relationships
\begin{subequations}
	\begin{align}
		\Omega^{2*}&=\Omega^{2},\label{eq:3.16a}\\
		\left\langle \delta\Psi_{-}|\mathbb{L}_{-}|\delta\Psi_{-}\right\rangle &=\Omega\Omega^{*}\left\langle \delta\Psi_{+}|\delta\Psi_{+}\right\rangle,\label{eq:3.16b}
	\end{align}
\end{subequations}
where the operation $\left\langle \cdot|\cdot\right\rangle $ is defined as the full-space integration $\left\langle \Psi_{1}|\Psi_{2}\right\rangle=\int \Psi^{*}_{1}\Psi_{2}dx^{3}$.
For the fundamental $Q$-ball with a monotonic radial profile function, the left side of (\ref{eq:3.16b}) is shown to be negative \cite{Panin:2016ooo}, which can not be equal to the positive definite term on the right side. Therefore, the relation (\ref{eq:3.16a}) needs to be fulfilled, indicating a real or purely imaginary eigenfrequency.
On the other hand, for the excited $Q$-balls with nodes in the radial profile function, the above conclusion is invalid, so the mode with a complex eigenfrequency is expected to appear, which should satisfy the relation (\ref{eq:3.16b}).

%%%%%%%%%%%%%%%%%%
\begin{figure}
	\centering
	\includegraphics[width=.49\linewidth]{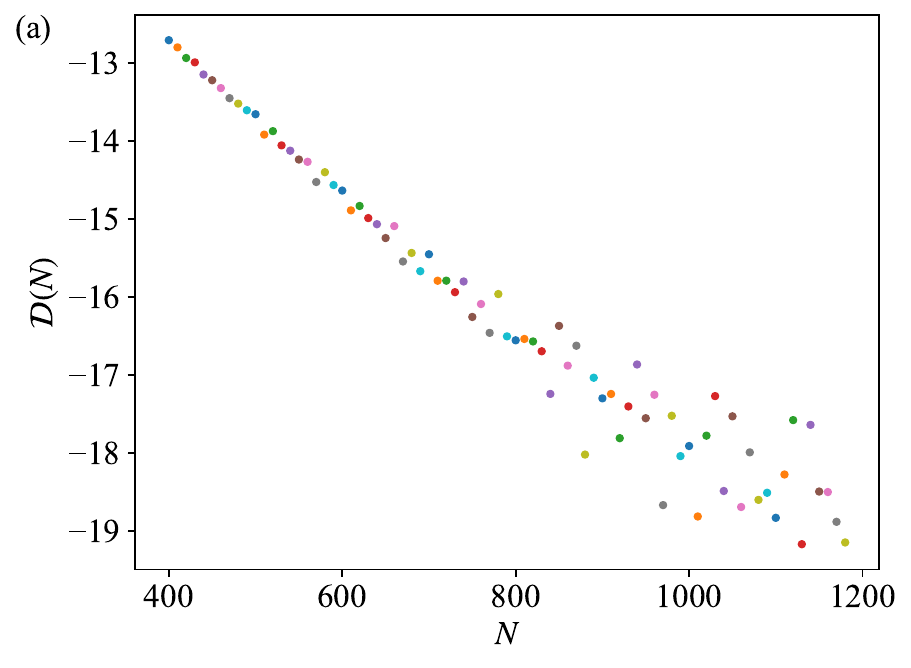}
	\includegraphics[width=.49\linewidth]{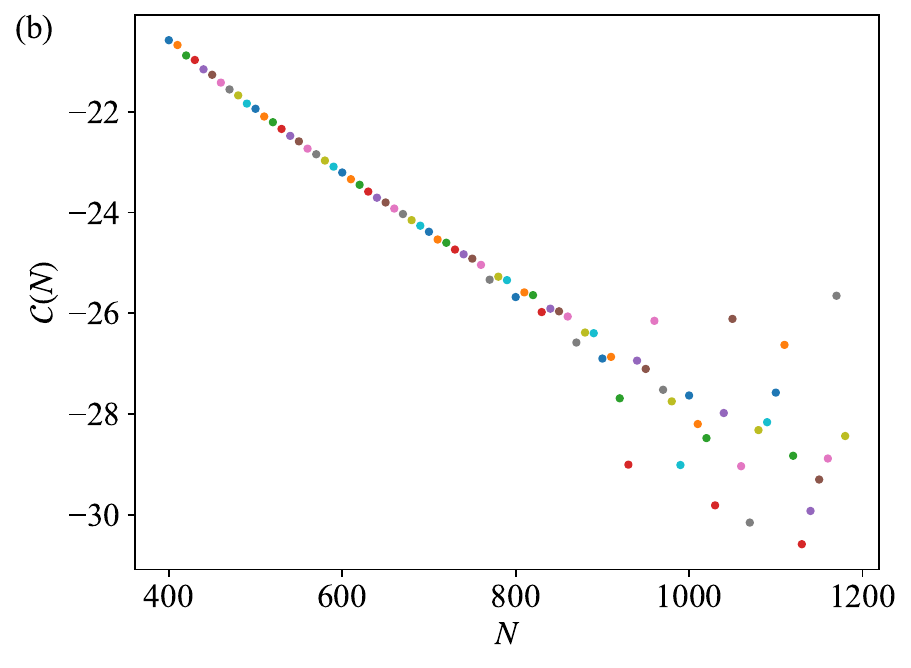}
	\caption{Numerical convergence and accuracy. The difference function $\mathcal{D}(N)$ (a) and the error function $\mathcal{C}(N)$ (b) change with the number of spectral grids $N$. The example in the figure is the mode with a complex eigenfrequency for an excited $Q$-ball.}
	\label{fig:5-6}
\end{figure}
%%%%%%%%%%%%%%%%%%

In order to evaluate the accuracy of the obtained numerical results, we define the following difference function over successive calculations
\begin{equation}
	\mathcal{D}(N)=\ln\left(|\Omega(N+\Delta N)-\Omega(N)|/\Delta N\right),
\end{equation}
which characterizes how the approximation of the eigenfrequency approaches the true value as the order of the polynomial basis increases, which is equivalent to the number of spectral grids $N$.
$\Delta N$ represents the interval of the number of increasing spectral grids, which is set to $10$ in this paper.
On the other hand, for the excited $Q$-balls, as mentioned above, the mode with a complex eigenfrequency needs to satisfy the additional constraint (\ref{eq:3.16b}).
Therefore, we use the following absolute error function 
\begin{equation}
	\mathcal{C}(N)=\ln|\left\langle \delta\Psi_{-}|\mathbb{L}_{-}|\delta\Psi_{-}\right\rangle (N)-\Omega\Omega^{*}\left\langle \delta\Psi_{+}|\delta\Psi_{+}\right\rangle (N)|,
\end{equation}
to measure the violation of such a constraint.
Taking the mode with a complex eigenfrequency of an excited $Q$-ball as an example, Figure \ref{fig:5-6} shows the convergence and accuracy of the numerical method.
Both the difference function $\mathcal{D}(N)$ and the error function $\mathcal{C}(N)$ decrease linearly with the increase of the number of spectral grids $N$, until a given optimal truncation order is reached, after which they exhibit a random distribution under an upper limit due to the dominance of possible sources of numerical inaccuracies such as computational errors.
Such results show that the obtained eigenfrequency converges exponentially to the true value, and the error also decreases exponentially, demonstrating the exponential convergence of the numerical method, which is exactly the convergence accuracy of the spectral method.
The same exponential convergence is also observed for modes with purely real or purely imaginary eigenfrequencies.

%=======================================================================

%=======================================================================
\section{Linear dynamical property}\label{sec:Ds}
Using the above numerical method, in this section we systematically reveal the dynamical stability of $Q$-balls from the fundamental state to the excited states.

\subsection{The fundamental $Q$-ball}
For the fundamental $Q$-ball, based on the relation (\ref{eq:3.16a}), one can label the eigenfrequencies as $\Omega^{2}_{0}<\cdots<\Omega^{2}_{\lambda}<\cdots$, where the mode with $\lambda=0$ is known as the fundamental mode and modes with $\lambda>0$ are excited modes.
On the other hand, due to the classical stability criterion, it can be predicted that all $Q$-balls can only possess zero modes and oscillation modes without imaginary parts, namely $\Omega^{2}_{0}\geq 0$, indicating the dynamical stability of the system.

%%%%%%%%%%%%%%%%%%
\begin{figure}
	\centering
	\includegraphics[width=.49\linewidth]{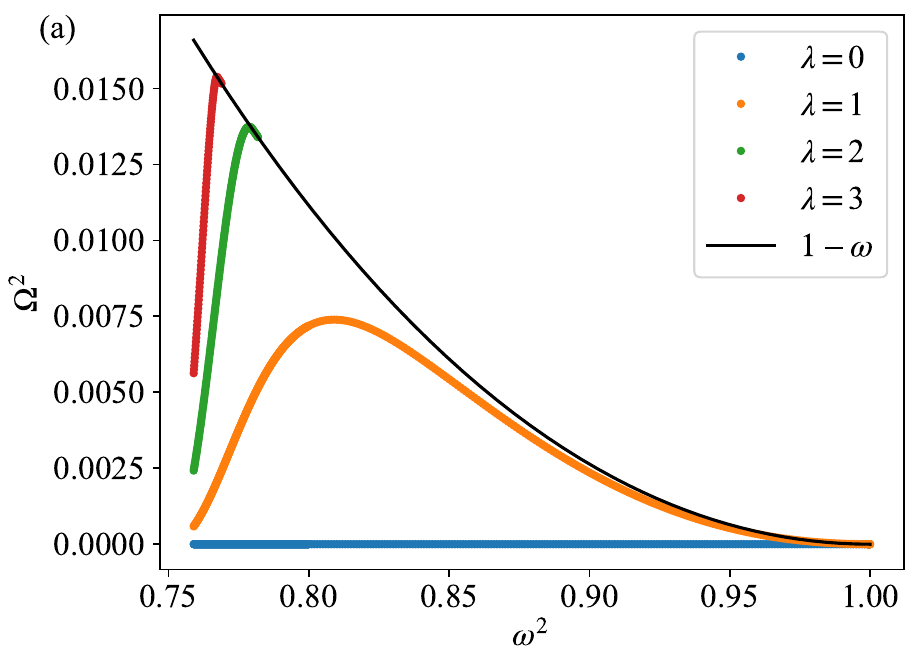}
	\includegraphics[width=.49\linewidth]{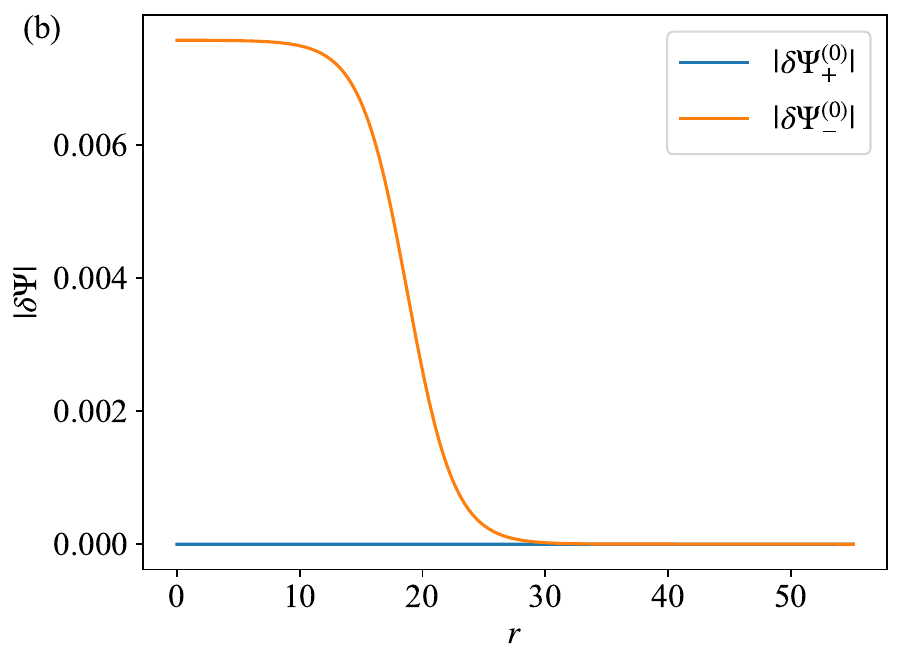}
	\includegraphics[width=.49\linewidth]{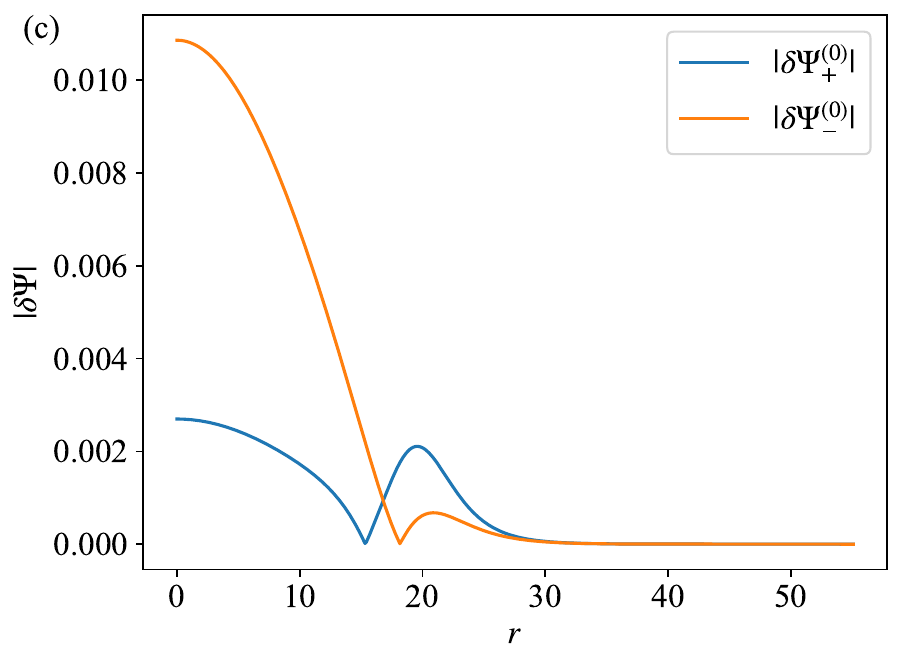}
	\includegraphics[width=.49\linewidth]{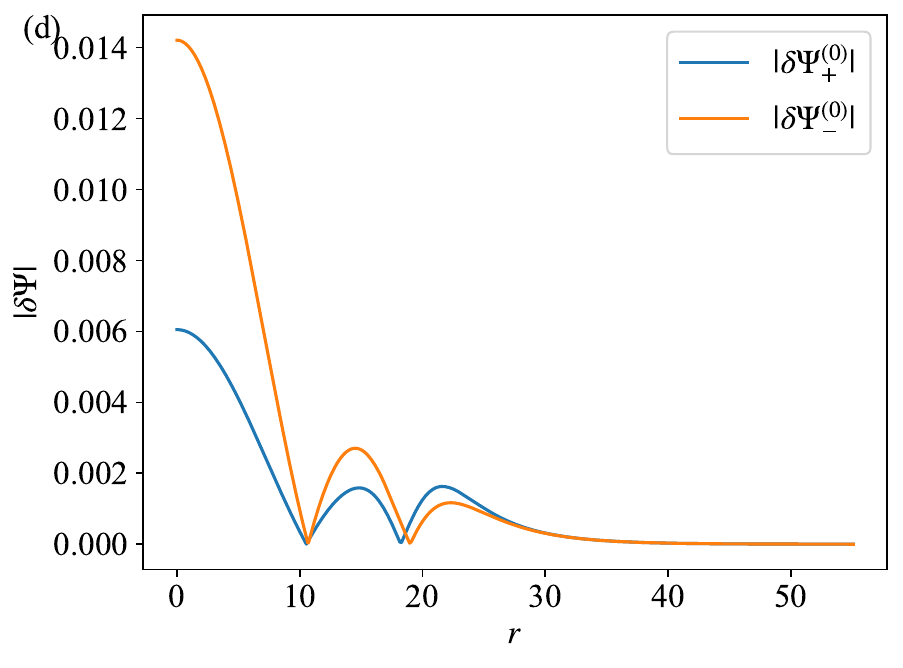}
	\caption{(a) The discrete spectrum with $l=0$. (b, c, d) The eigenstates of the zero mode $\lambda=0$ and the first two oscillation modes $\lambda=1,2$ of the $Q$-ball with frequency $\omega^{2}=0.77$.}
	\label{fig:7-10}
\end{figure}
%%%%%%%%%%%%%%%%%%

%%%%%%%%%%%%%%%%%%
\begin{figure}
	\centering
	\includegraphics[width=.49\linewidth]{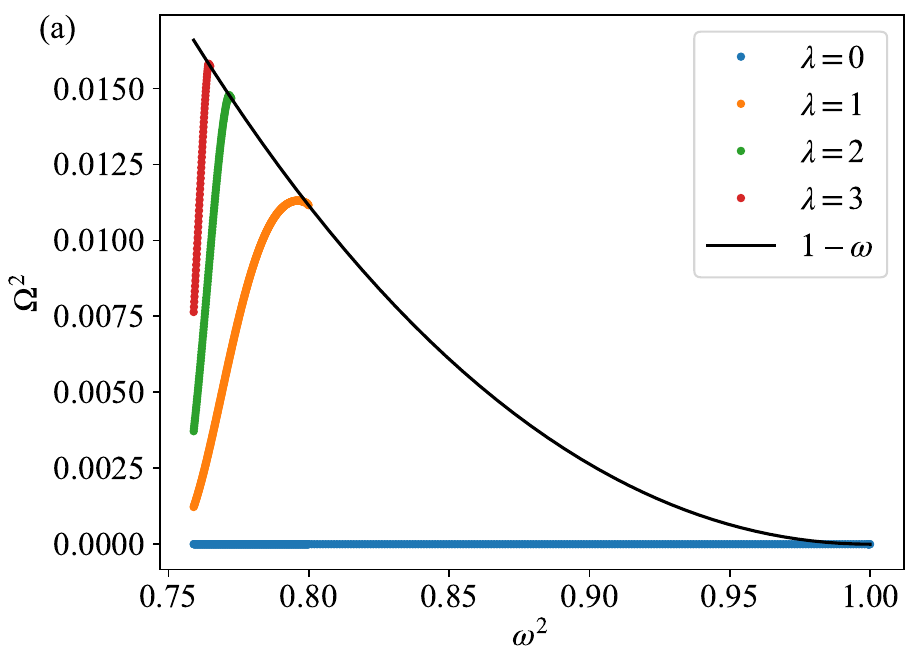}
	\includegraphics[width=.49\linewidth]{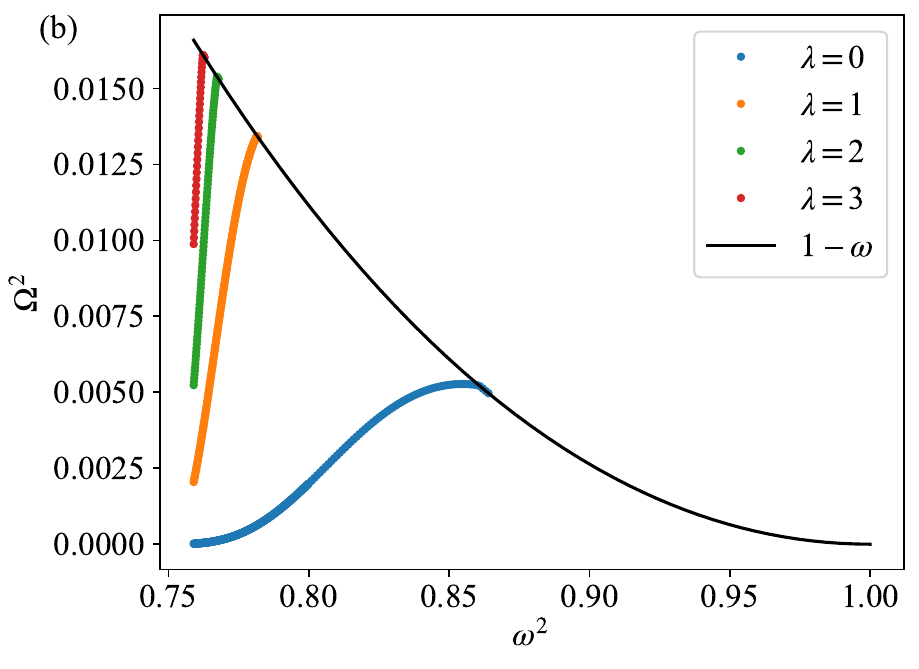}
	\includegraphics[width=.49\linewidth]{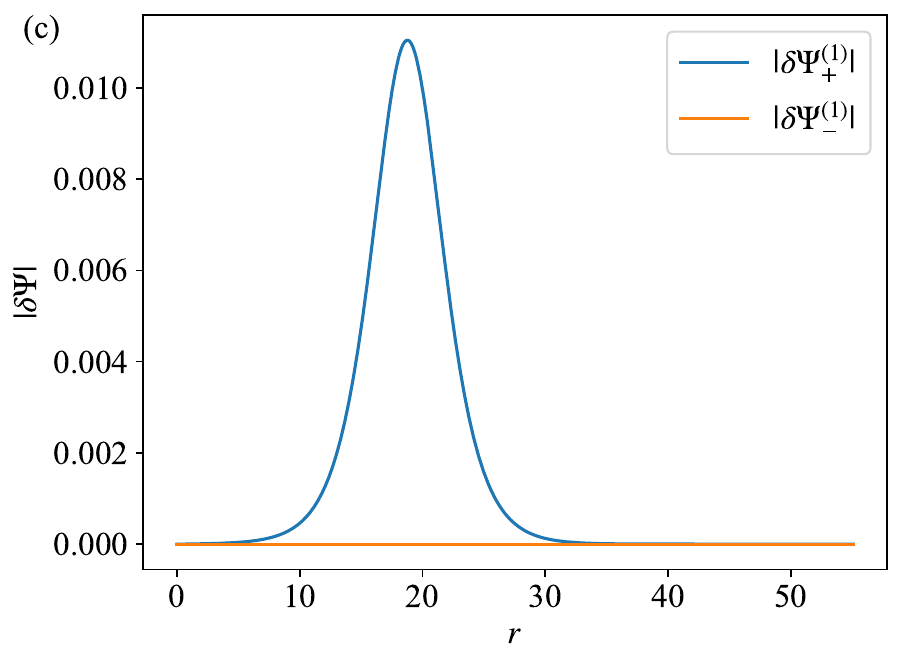}
	\includegraphics[width=.49\linewidth]{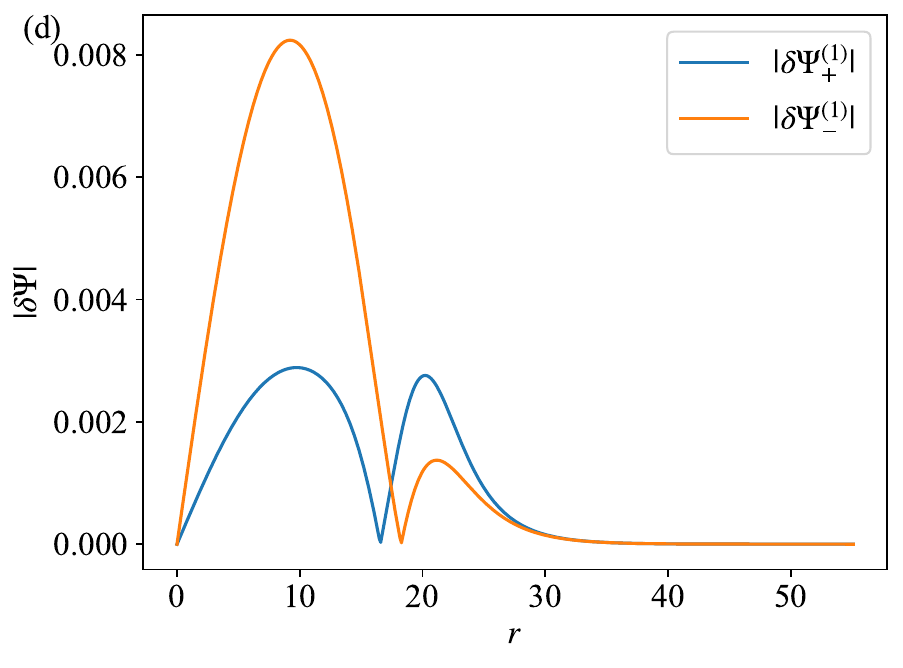}
	\caption{The discrete spectra with $l=1$ (a) and $l=2$ (b). (c, d) The eigenstates of the zero mode $\lambda=0$ and the first oscillation mode $\lambda=1$ of the $Q$-ball with frequency $\omega^{2}=0.77$ for the perturbation with $l=1$. }
	\label{fig:11-14}
\end{figure}
%%%%%%%%%%%%%%%%%%

The numerical results for the perturbation of spherical symmetry with $l=0$ are shown in Figure \ref{fig:7-10}.
The eigenfrequency is discretely distributed within the interval $[0,1-\omega)$ along the real axis, demonstrating the dynamical stability of $Q$-balls.
In the thick-wall region, the $Q$-ball possesses a single oscillation mode $\lambda=1$.
Approaching the thin-wall limit, more oscillation modes (only the first two are shown in the figure $\lambda=2,3$) break in from the outer boundary of the interval, and there is a tendency to converge to zero.
Such a convergence behavior in the thin-wall limit can be proved by the analytical approximation method \cite{Kovtun:2018jae}. 
The value of the integer $\lambda$ corresponds to the number of nodes in the radial profile of the eigenstate $\delta\Psi_{\pm}$, analogous to the scenario of the radial profile function of excited $Q$-balls with the number of nodes $n$.
In addition, all $Q$-balls have a zero mode $\lambda=0$ with an eigenstate $\delta\Psi^{(0)}_{\pm}=\{0,C\phi\}$ associated with its radial profile function, where $C$ is a proportionality constant. The existence of such a zero mode can be analytically proved. More precisely, substituting $\Omega=0$ into (\ref{eq:3.15}), the perturbation equations degenerate into the following form
\begin{equation}
	\begin{pmatrix}
		\mathbb{L}^{(l)}_{+}&0\\
		0&\mathbb{L}^{(l)}_{-}
	\end{pmatrix}
	\begin{pmatrix}
		\delta\Psi^{(l)}_{+}\\
		\delta\Psi^{(l)}_{-}
	\end{pmatrix}=0.\label{eq:4.1}
\end{equation}
Since the stationary equation (\ref{eq:2.12}) satisfied by the radial profile function of $Q$-balls is equivalent to $\mathbb{L}^{(0)}_{-}\phi=0$, the function group $\{0,C\phi\}$ is the solution to the above equation with $l=0$.

For the perturbation of non-spherical symmetry with $l=1$, the situation is similar to that for the spherically symmetric perturbation, as shown in figure \ref{fig:11-14}.
There are two main differences.
One is that there is no oscillation mode in the thick-wall region, and the other is that the value of the radial profile function of the eigenstate at the origin is zero due to the asymptotic behavior (\ref{eq:3.5a}), namely $\delta\Psi^{(l>0)}_{\pm}(0)=0$, similar to the scenario of the radial profile of spinning $Q$-balls \cite{Volkov:2002aj}.
Since the derivative of the stationary equation (\ref{eq:2.12}) with respect to the radial coordinate is equivalent to $\mathbb{L}^{(1)}_{+}\frac{d\phi}{dr}=0$, the eigenstate of the zero mode at this time is $\delta\Psi^{(1)}_{\pm}=\{C\frac{d\phi}{dr},0\}$, which satisfies the equation (\ref{eq:4.1}) with $l=1$.
For higher order perturbations with $l>1$, $Q$-balls will no longer have a zero mode, namely $\Omega^{2}_{0}>0$.
Such results indicate that the classical stability criterion is applicable to the fundamental $Q$-ball, which is consistent with the conclusions obtained from the linear stability analysis.

\subsection{The first excited $Q$-ball}
For the excited $Q$-balls, the existence of nodes in the radial profile function makes many existing conclusions no longer applicable, such as the failure of the stability criterion and the existence of modes with complex eigenfrequencies.
On the other hand, since the excited $Q$-balls always possess higher energy, as shown in Figure \ref{fig:3-4}, it can be believed that they are dynamically unstable and can spontaneously degenerate to the ground state with lower energy, that is, the fundamental $Q$-ball.
However, the numerical results show that this is not always the case.

%%%%%%%%%%%%%%%%%%
\begin{figure}
	\centering
	\includegraphics[width=.49\linewidth]{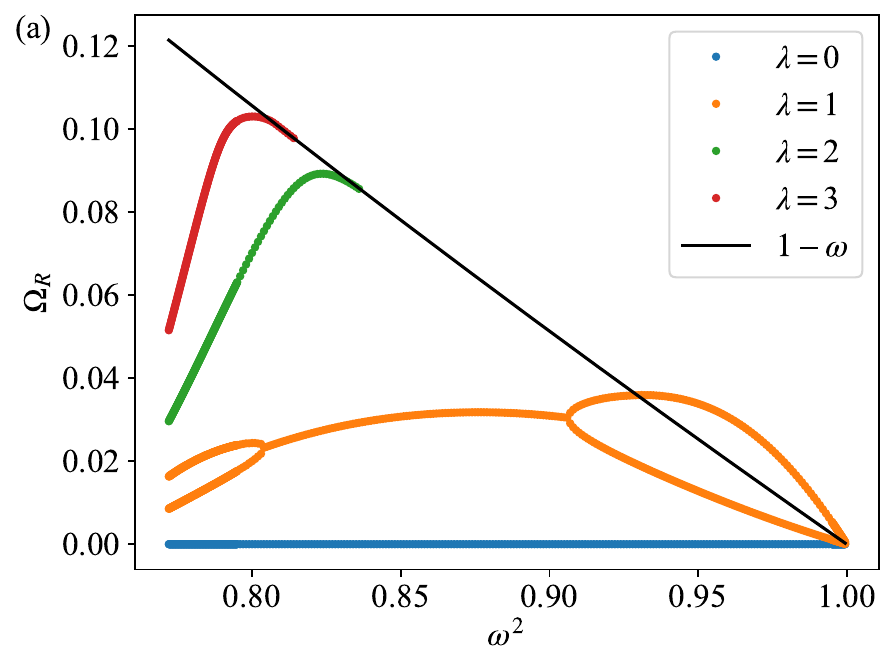}
	\includegraphics[width=.49\linewidth]{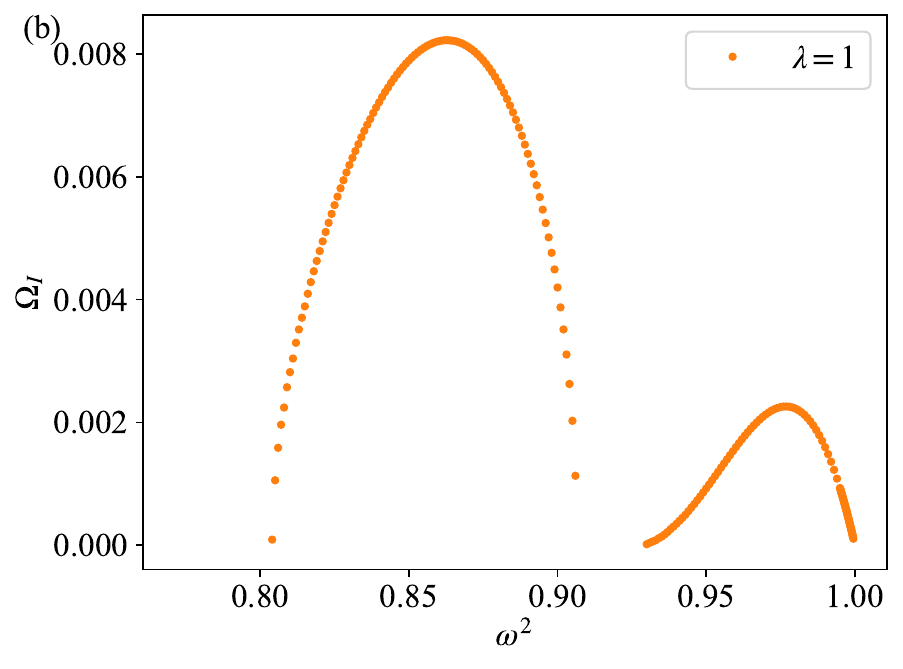}
	\caption{The discrete spectrum with $l=0$ for the first excited $Q$-ball, with the real part of modes (a) and the imaginary part of the unstable mode (b).}
	\label{fig:15-16}
\end{figure}
%%%%%%%%%%%%%%%%%%

\begin{figure}
	\centering
	\includegraphics[width=.49\linewidth]{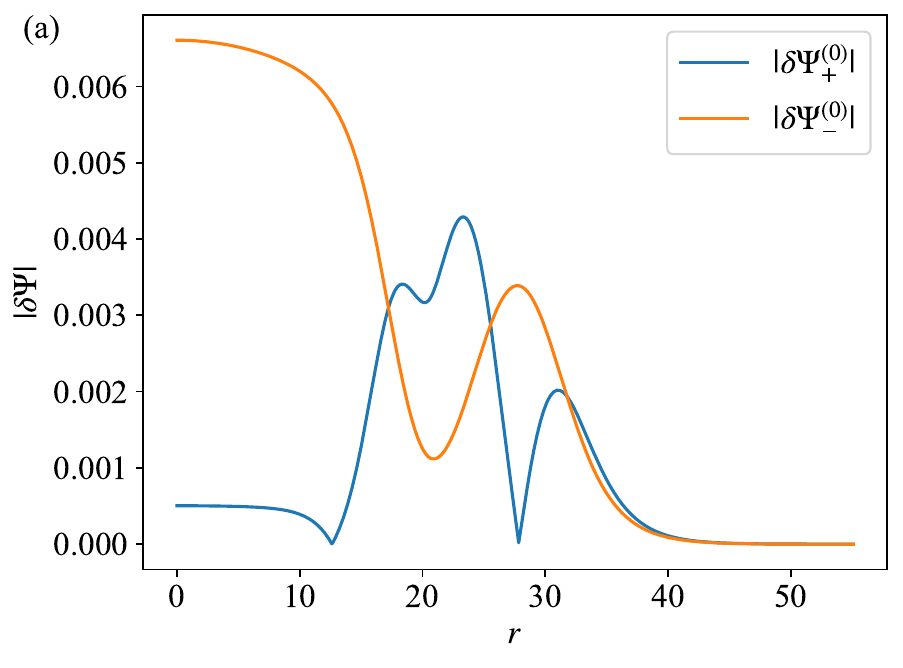}
	\includegraphics[width=.49\linewidth]{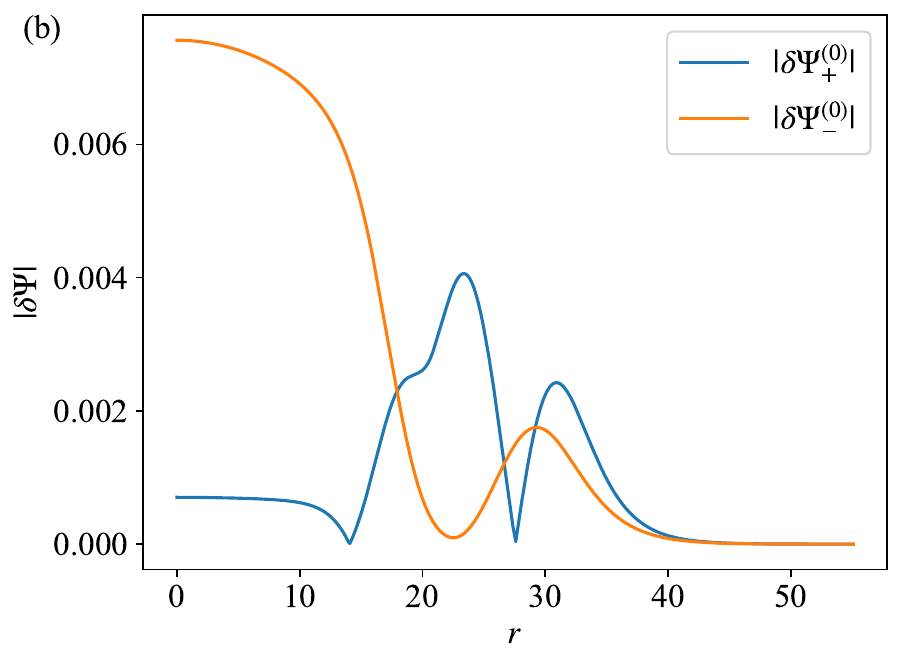}
	\includegraphics[width=.49\linewidth]{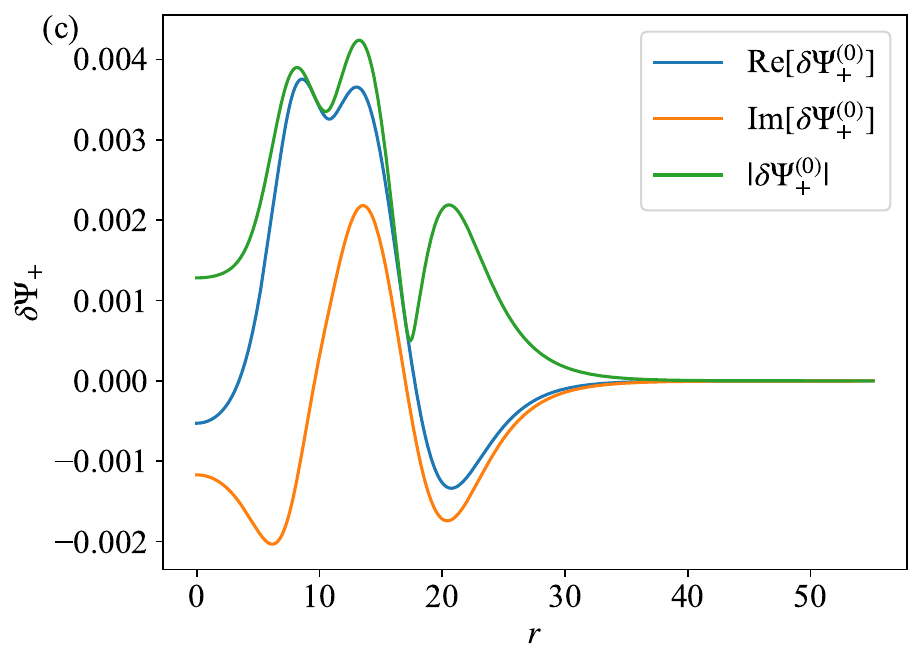}
	\includegraphics[width=.49\linewidth]{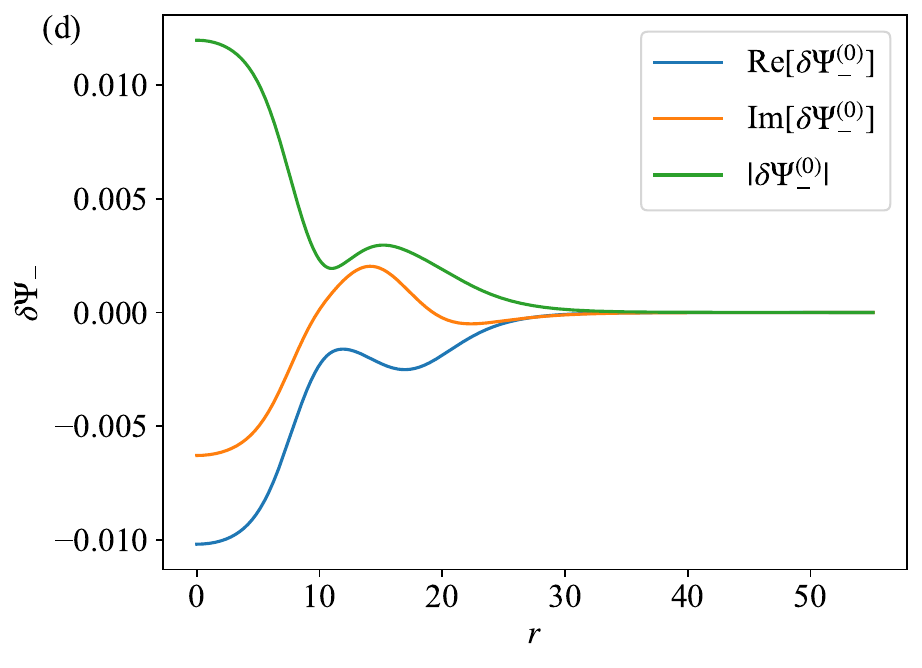}
	\includegraphics[width=.49\linewidth]{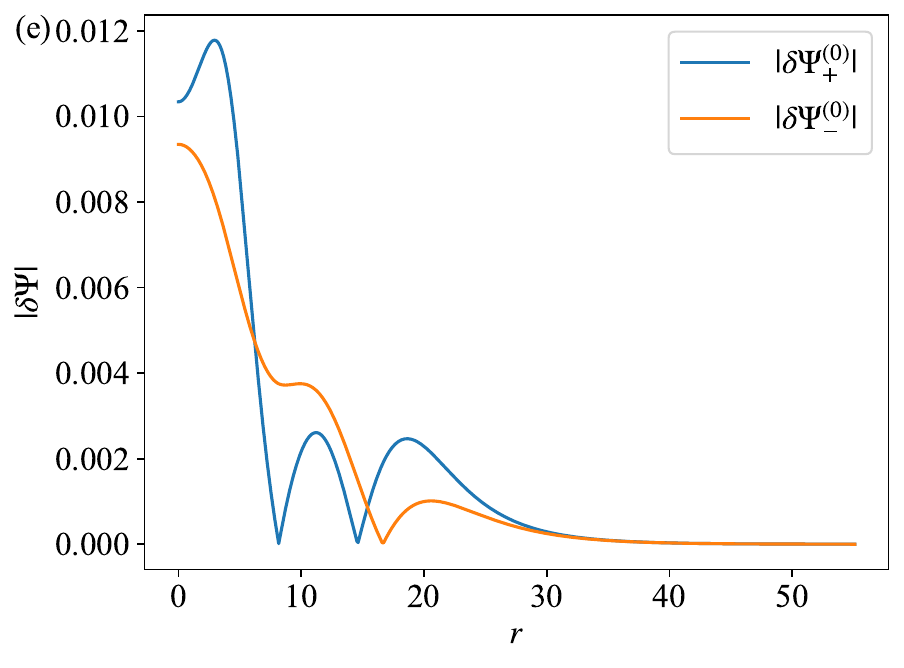}
	\includegraphics[width=.49\linewidth]{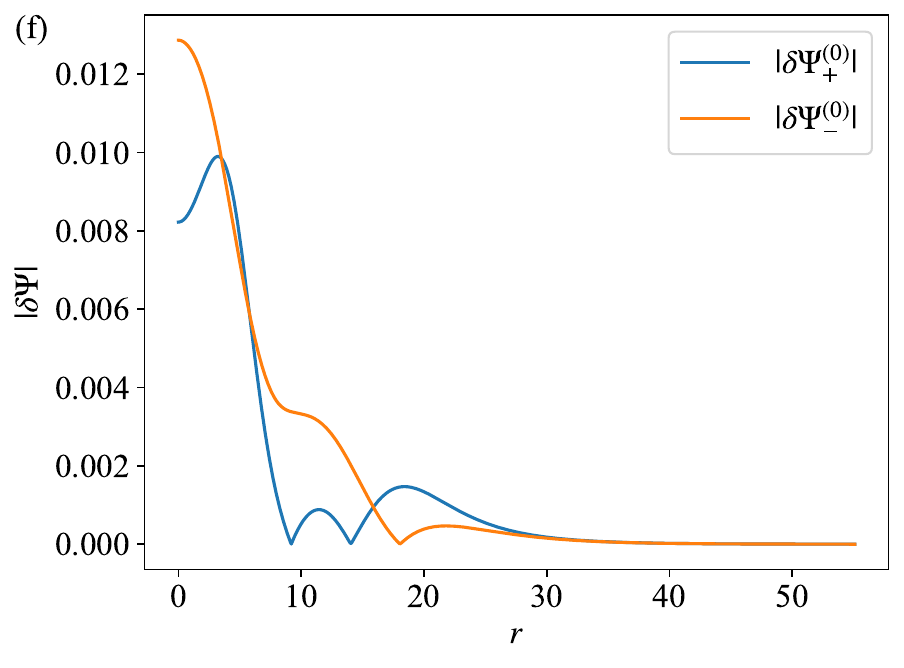}
	\caption{(a, b) The eigenstates of the small and large oscillation modes for the $Q$-ball with frequency $\omega^{2}=0.8$. (c, d) The components $\delta\Psi^{(0)}_{+}$ and $\delta\Psi^{(0)}_{-}$ of the eigenstate of the complex mode for the $Q$-ball with frequency $\omega^{2}=0.85$. (e, f) The eigenstates of the small and large oscillation modes for the $Q$-ball with frequency $\omega^{2}=0.907$.}
	\label{fig:17-22}
\end{figure}

Figure \ref{fig:15-16} shows the discrete spectrum of the first excited $Q$-ball for the perturbation of spherical symmetry with $l=0$.
Interestingly, near the thin-wall limit, similar to the scenario of fundamental $Q$-ball, the system exhibits dynamical stability.
There exists a zero mode $\lambda=0$ and some oscillation modes $\lambda=2,3$, the number of which gradually increases near the thin-wall limit.
The difference is that there is a pair of dual oscillation modes $\lambda=1$.
In general, the eigenstate of the oscillation mode with a larger eigenfrequency has more nodes.
Moreover, in the mode queue, the number of nodes in the eigenstates of adjacent oscillation modes differs by one, such as the results obtained above.
However, this is not always the case.
As demonstrated above, the eigenstate of the zero mode for the spherically symmetric perturbation is $\delta\Psi^{(0)}_{\pm}=\{0,C\phi\}$ with the radial profile function $\phi$ of the perturbed $Q$-ball.
Here the perturbed background is the first excited $Q$-ball, whose profile function has one node.
We have checked that the eigenstates of oscillation modes $\lambda=2,3$ have three and four nodes respectively, lacking the oscillation mode with two nodes.
As can be observed from Figure \ref{fig:17-22}, the eigenstates of the pair of dual oscillation modes $\lambda=1$ have similar structures, characteristically, of which the components $\delta\Psi^{(0)}_{+}$ have two nodes.
Therefore, it is reasonable to consider that such a pair of dual oscillation modes is derived from the lesion of the missing oscillation mode with two nodes. 
In the following, the one with the smaller eigenfrequency in a pair of dual oscillation modes is labeled as the small oscillation mode, and the other with the larger eigenfrequency is called the large oscillation mode.

%%%%%%%%%%%%%%%%%%
%%%%%%%%%%%%%%%%%%

The dynamical stability of first excited $Q$-balls near the thin-wall limit suggests that the common perception of excited states, which are dynamically unstable and can spontaneously degenerate to a ground state, does not always hold true in the case of $Q$-balls.
However, away from the thin-wall limit, things start to get better.
The dual oscillation modes in the mode pair $\lambda=1$ gradually approach each other and merge together at a critical excited $Q$-ball, forming an oscillation mode with an algebraic multiplicity of two, labeling the corresponding critical frequency as $\omega_{a}$.
Subsequently, such a degenerate oscillation mode further splits into a pair of complex modes with conjugate symmetry, in which the one with a positive imaginary part triggers the dynamical instability.
Such a transition from the oscillation mode to the complex mode marks the transition of the system from stability to instability.
Different from the real eigenstates of the zero mode and oscillation modes, the eigenstate of the complex mode is a combination of complex functions, as shown in Figure \ref{fig:17-22}, of which the imaginary parts of both components have two nodes.

Interestingly, the story is not over yet.
Moving further away from the thin-wall limit, the imaginary part of the complex mode, characterizing the growth rate of the dynamical instability, gradually increase until it reaches saturation.
After that, surprisingly, the dynamical instability is suppressed and finally disappears at a critical excited $Q$-ball, labeling the corresponding critical frequency as $\omega_{b}$.
At this time, the complex mode migrates to the real axis and coincides with its complex conjugate symmetric counterpart, forming an oscillation mode with an algebraic multiplicity of two again.
Subsequently, such a degenerate oscillation mode splits into a pair of dual oscillation modes, one of which migrates toward the zero mode, while the other one translates toward the interval boundary in the opposite direction.
Such a transition from the complex mode to the oscillation mode marks the transition of the system from instability to stability.
Similarly, the eigenstates of the small and large oscillation modes here show a high degree of similarity, manifested as two nodes in the components $\delta\Psi^{(0)}_{+}$, as shown in Figure \ref{fig:17-22}.
Differently, with the migration of these two oscillation modes, the two nodes in the eigenstate of the small oscillation mode gradually separate, while those of the large oscillation mode gradually approach.
Finally, for the case of large oscillation mode, the two nodes merge together and then evaporate.
The story continues.
Such a large oscillation mode will break through the interval boundary and transform into a complex mode, accompanied by the generation of the imaginary part of the eigenfrequency, indicating that the first excited $Q$-balls in the thick-wall region are also dynamically unstable to the spherically symmetric perturbation, although this dynamical instability is suppressed again as the thick-wall limit is approached.
Such results demonstrate the existence of excited $Q$-balls capable of resisting the spherically symmetric perturbation, similar to what happens in excited boson stars \cite{Sanchis-Gual:2021phr,Brito:2023fwr,Nambo:2024gvs}.

%%%%%%%%%%%%%%%%%%
\begin{figure}
	\centering
	\includegraphics[width=.49\linewidth]{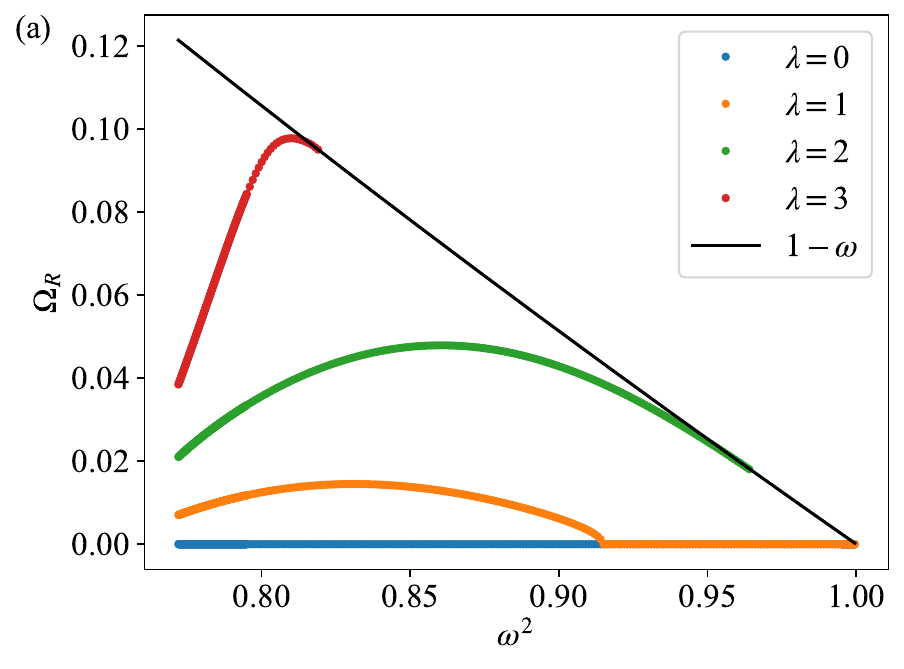}
	\includegraphics[width=.49\linewidth]{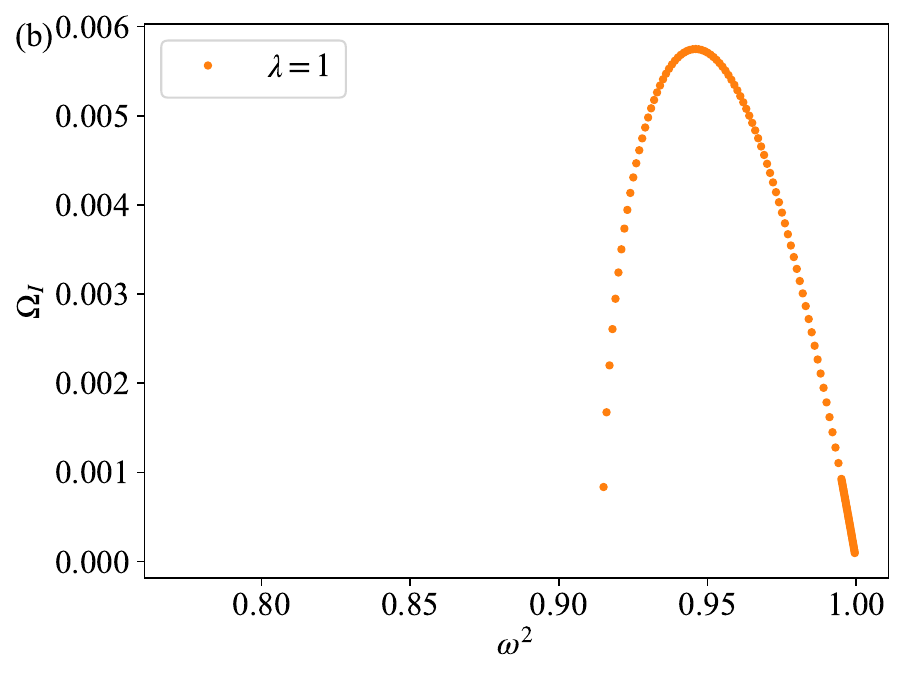}
	\includegraphics[width=.49\linewidth]{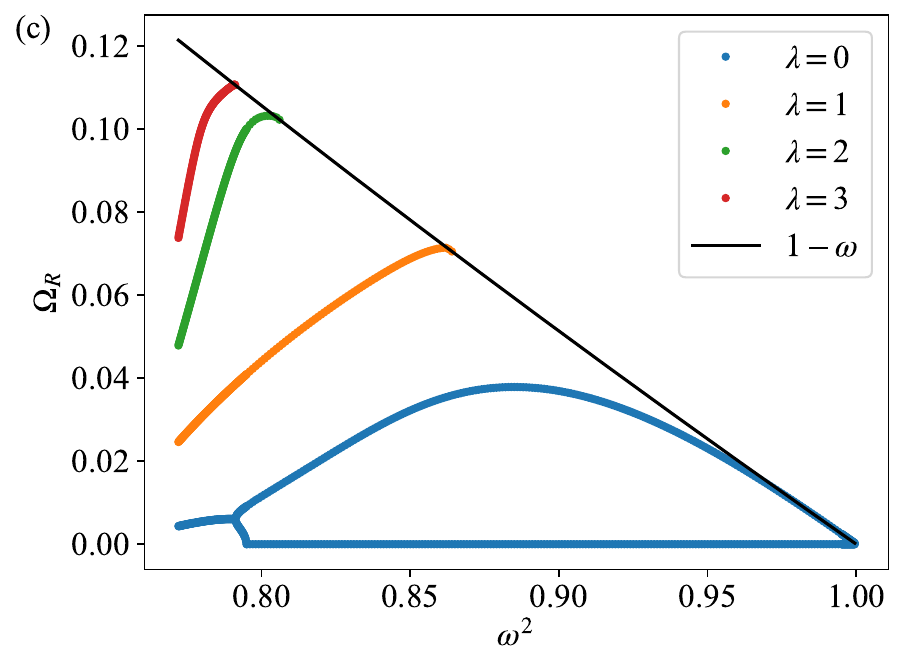}
	\includegraphics[width=.49\linewidth]{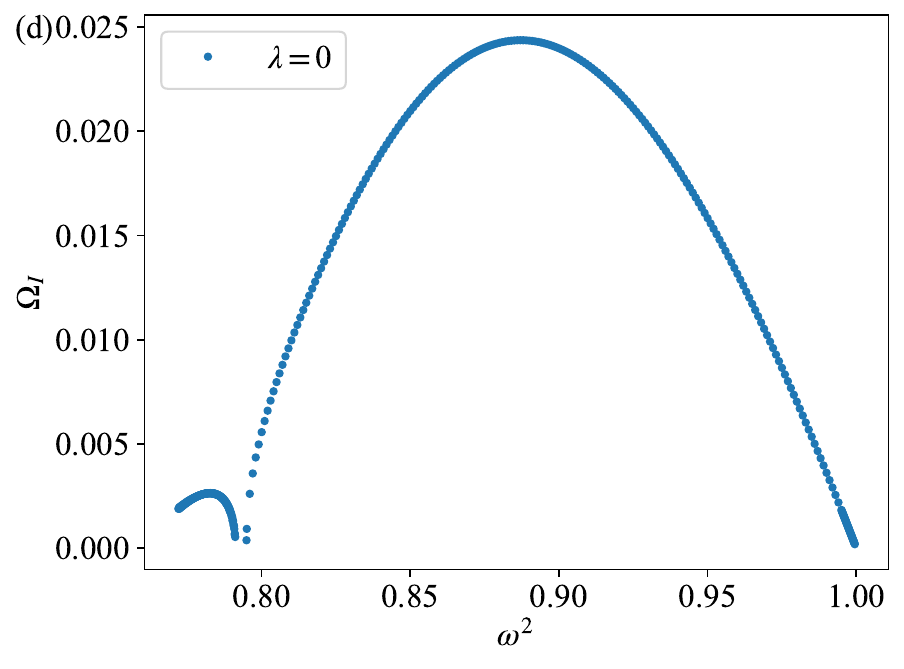}
	\includegraphics[width=.49\linewidth]{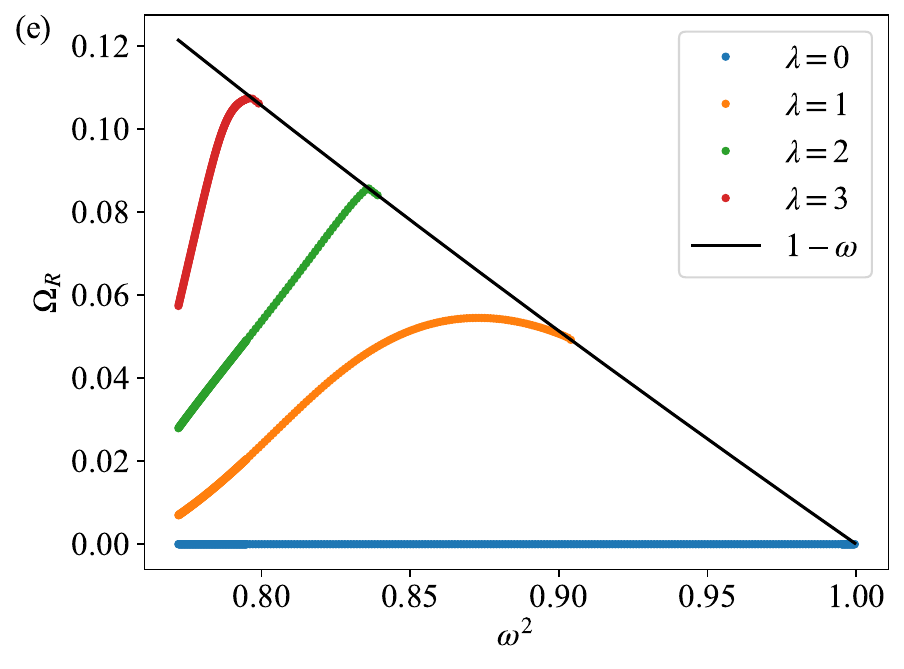}
	\includegraphics[width=.49\linewidth]{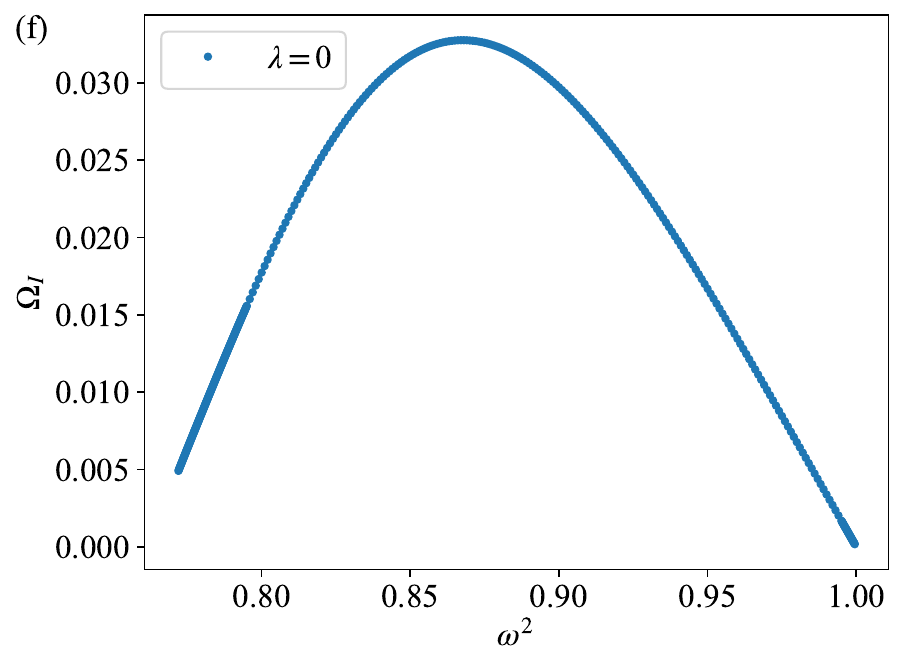}
	\caption{The discrete spectra with $l=1$ (upper panel), $l=2$ (middle panel) and $l=3$ (lower panel) for the first excited $Q$-ball, with the real part of modes (left column) and the imaginary part of the unstable mode (right column).}
	\label{fig:23-28}
\end{figure}
%%%%%%%%%%%%%%%%%%

A natural question is whether the excited $Q$-balls that are resistant to the spherically symmetric perturbation can also maintain dynamically stable under non-spherically symmetric perturbations.
Figure \ref{fig:23-28} shows the discrete spectra of the first excited $Q$-ball for the perturbations of non-spherical symmetry with $l=1$, $l=2$ and $l=3$.
For the perturbation with $l=1$, there is a zero mode $\lambda=0$ and some oscillation modes $\lambda=2,3$.
The eigenstate of the zero mode is related to the derivative of the profile function of the first excited $Q$-ball with respective to the radial coordinate $\frac{d\phi}{dr}$, with two nodes including one at the origin.
On the other hand, we observe three and four nodes in the eigenstates of oscillation modes $\lambda=2,3$, respectively.
Therefore, these three modes should be ordered sequentially in the queue $\Omega^{2}_{0}<\Omega^{2}_{2}<\Omega^{2}_{3}<\cdots$.
Surprisingly, there is an additional mode $\lambda=1$, which behaves as an oscillation mode in the thin-wall region and transforms into a purely imaginary mode in the thick-wall region. 
Therefore, there exists a critical excited $Q$-ball with a zero mode with algebraic multiplicity of two, labeling the corresponding critical frequency as $\omega_{c}$.
Such results indicate that the first excited $Q$-balls in the thin-wall region $\omega<\omega_{c}$ are able to resist the non-spherically symmetric perturbation with $l=1$.
Fortunately, this critical frequency here lies within the stability interval in the case of spherically symmetric perturbation, indicating that the first excited $Q$-balls in the regions $\omega<\omega_{a}$ and $\omega_{b}<\omega<\omega_{c}$ are immune to both the perturbations of spherical symmetry and non-spherical symmetry with $l=1$.

However, the situation changes essentially for perturbations with high-order spherical harmonics.
For the case of $l=2$, near the thin-wall limit, there is a complex mode $\lambda=0$ \footnote{Since the profile function of the $Q$-ball at the thin-wall limit tends to a discontinuous function, which is difficult to approximate with a combination of finite-order polynomials, it is difficult for numerical methods to capture the behavior in such a case. Therefore, we cannot determine the situation near the thin-wall limit, for example, whether the complex mode $\lambda=0$ is generated by the collision of a pair of dual oscillation modes, similar to what happens in the case of spherically symmetric perturbation.} that dominates the dynamical instability of the system.
Away from the thin-wall limit, similar to the situation of spherically symmetric perturbation, this complex mode with its conjugate symmetric counterpart will transform into a pair of dual oscillation modes at a critical excited $Q$-ball, labeling the corresponding critical frequency as $\omega_{d}$.
Subsequently, the large oscillation mode migrates to the interval boundary, while the small oscillation mode quickly approaches the origin and becomes a zero mode. 
Such a mode will further cross the origin and continue to climb along the imaginary axis, transforming into a purely imaginary mode and dominating the dynamical instability of the system.
This instability extends to the thick-wall limit, showing that only the first excited $Q$-balls in a very small region close to the thin-wall limit are able to resist the perturbation with $l=2$.
Since the stability region here only overlaps with the above stability interval $\omega<\omega_{a}$, one can conclude that for a composite perturbation up to $l=2$, only the first excited $Q$-balls in the region $\omega_{d}<\omega<\omega_{a}$ exhibit dynamical stability, which is further eliminated by a purely imaginary mode in the case of perturbation with $l=3$.

Based on the above numerical results, for the first excited $Q$-ball, one can draw the following conclusions:
\begin{itemize}
	\item The classical stability criterion is no longer applicable.
	\item There is a dynamical instability that preserves the spherical symmetry, dominated by a complex mode.
	\item There are dynamical instabilities that break the spherical symmetry, dominated by a complex mode or a purely imaginary mode.
	\item There exist some configurations capable of resisting perturbations with low-order spherical harmonics.
	\item The oscillation mode can transform into the complex mode or the imaginary mode, and vice versa, marking a transition in the dynamical properties of the system.
\end{itemize}
The above properties indicate that the excited $Q$-ball will exhibit rich dynamical behaviors, manifested in the diversity of the final states of evolution.
In general, an excited state is expected to spontaneously degenerate to a ground state with lower energy, manifesting as a dynamical transition that releases energy.
For $Q$-balls here, since the fundamental $Q$-ball possesses lower energy, it is naturally considered a candidate for the final state of the excited $Q$-ball.
However, due to the existence of dynamical instability that destroys the spherical symmetry of the system, there may be multiple evolution paths. 
The basis for making such a conjecture is that the fundamental $Q$-ball is spherically symmetric and therefore cannot exist in a dynamical process with broken spatial symmetry.
Based on the above speculation, in general, there may be two scenarios.
For an excited $Q$-ball unstable to the spherically symmetric perturbation, a single fundamental $Q$-ball, a combination of several fundamental $Q$-balls, and a free particle are all candidates for the final state, depending on the geometry of perturbations.
However, for an excited $Q$-ball capable of resisting the spherically symmetric perturbation, the configuration can only be torn apart by non-spherically symmetric perturbations and evolve into several fundamental $Q$-balls or a free particle, or remain unchanged under the spherically symmetric perturbation.
These dynamical processes require further verification by nonlinear numerical simulations.

\subsection{The second excited $Q$-ball}
\begin{figure}
	\centering
	\includegraphics[width=.49\linewidth]{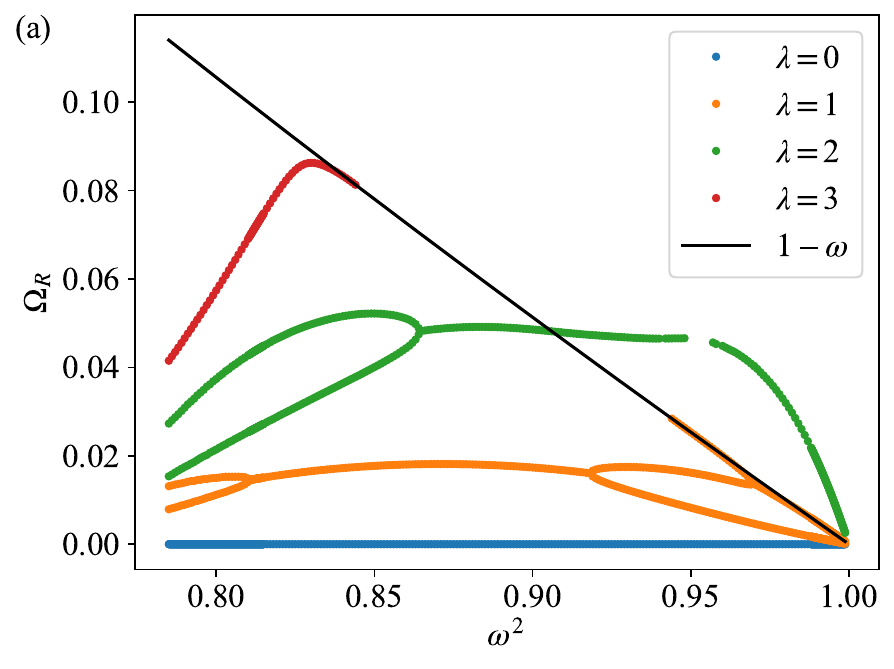}
	\includegraphics[width=.49\linewidth]{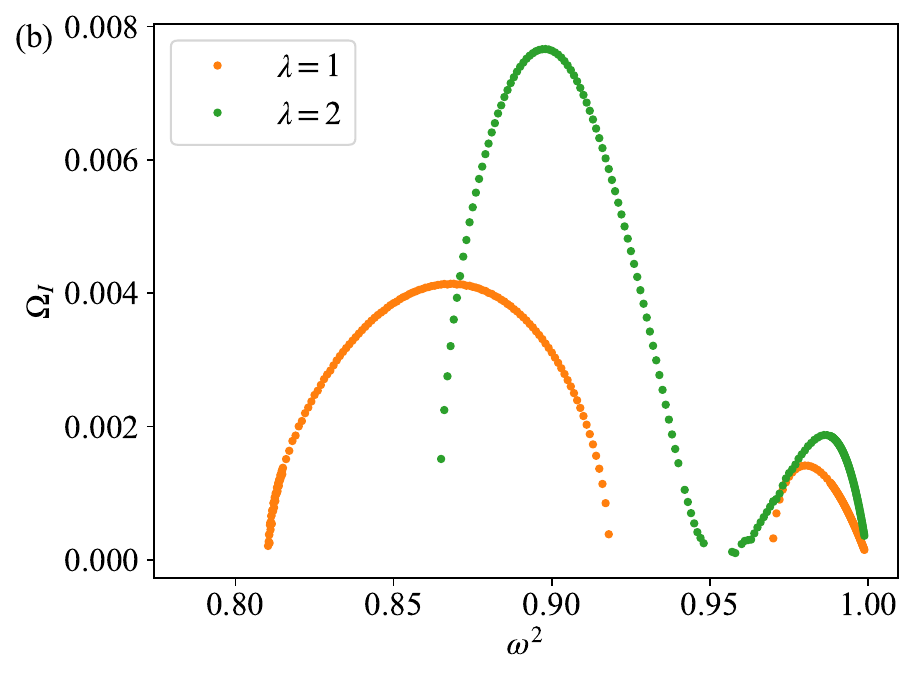}
	\includegraphics[width=.49\linewidth]{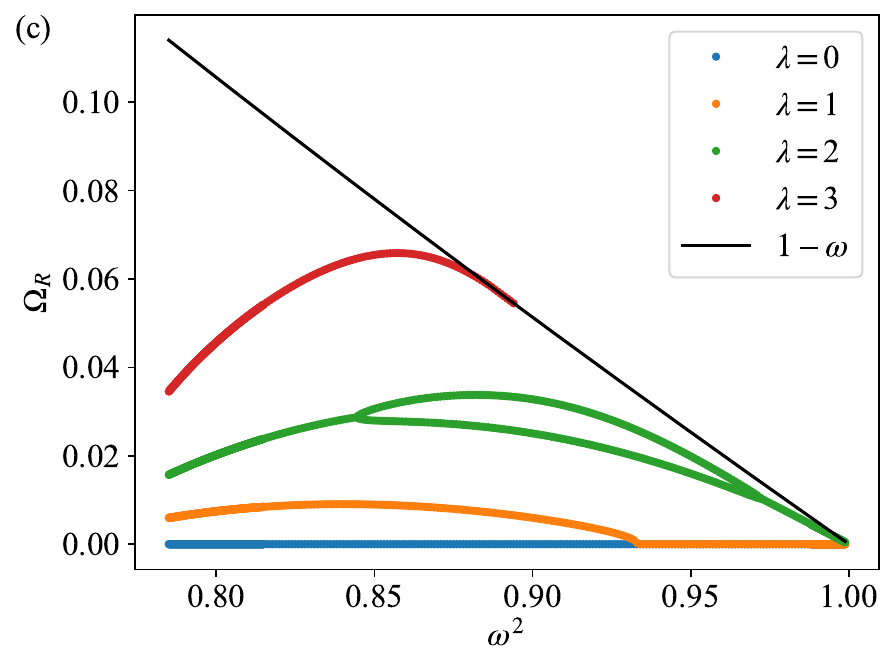}
	\includegraphics[width=.49\linewidth]{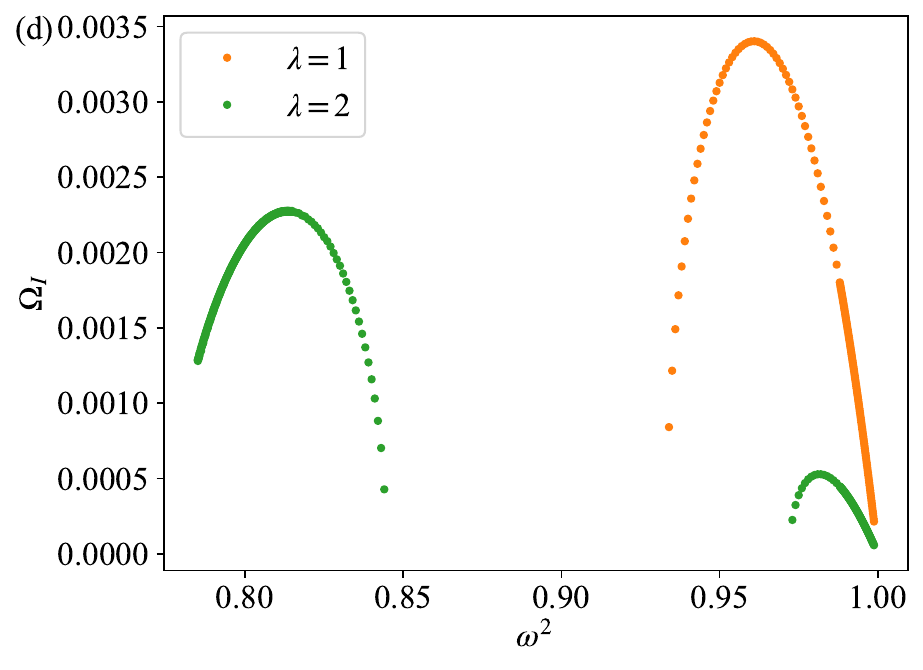}
	\includegraphics[width=.49\linewidth]{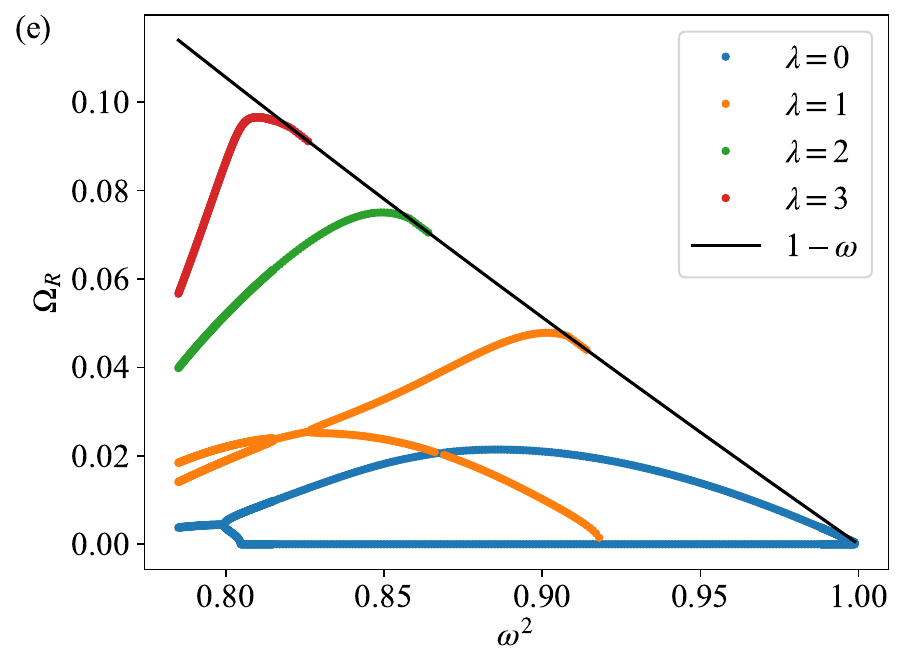}
	\includegraphics[width=.49\linewidth]{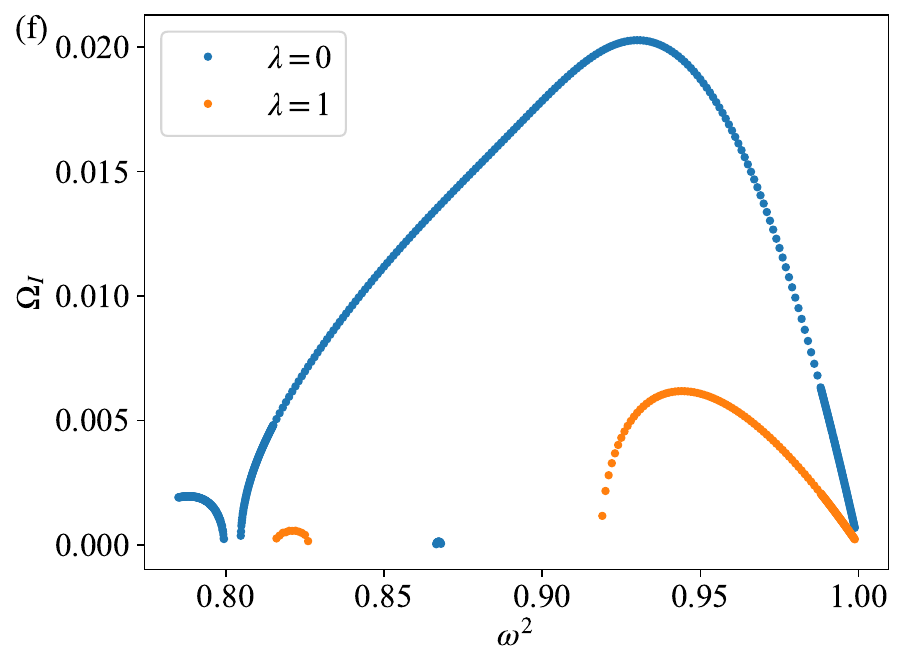}
	\caption{The discrete spectra with $l=0$ (upper panel), $l=1$ (middle panel) and $l=2$ (lower panel) for the second excited $Q$-ball, with the real part of modes (left column) and the imaginary part of the unstable mode (right column).}
	\label{fig:29-34}
\end{figure}

The results in the previous subsection confirm the existence of first excited $Q$-balls capable of resisting perturbations with low-order spherical harmonics.
In this subsection, we further reveal the linear dynamical properties of the second excited $Q$-ball. 
The conclusions drawn are similar to those above but with more complicated phenomena, as shown in Figure \ref{fig:29-34}.

%%%%%%%%%%%%%%%%%%
%%%%%%%%%%%%%%%%%%

For the perturbation of spherical symmetry with $l=0$, near the thin-wall limit, the system also exhibits dynamical stability, similar to the case of the first excited $Q$-ball.
The difference is that there are two oscillation modes that have undergone the lesion, generating two pairs of dual oscillation modes $\lambda=1,2$.
Among them, the migration behavior of the first mode pair $\lambda=1$ is similar to that in the case of the first excited $Q$-ball, undergoing two transitions.
After the intermediate process of transitioning to a complex mode pair that marks dynamical instability, it finally transforms into a pair of dual oscillation modes again.
In the case of the first excited $Q$-ball, the resulting large oscillation mode will break through the interval boundary and will transform into a complex mode.
Differently, for the second excited $Q$-ball here, as the large oscillation mode approaches the interval boundary, an additional oscillation mode suddenly invades from outside the interval boundary and collides with the coming large oscillation mode, forming a degenerate oscillation mode.
Such a collision behavior eventually leads to the transition to a pair of complex modes.
On the other hand, for the second mode pair $\lambda=2$, only one transition occurs, resulting in the emergence of a complex mode with a larger real part.
As the imaginary part grows, this complex mode gradually suppresses the previous one and dominates the dynamical instability of the system.
After crossing the interval boundary, it converges to the real axis~\footnote{Since the real part is outside the interval $[0,1-\omega)$, we cannot determine the exact details of what is happening here.} and then reappears near the thick-wall limit, leaving a stability interval.
In summary, as shown in Figure \ref{fig:29-34}(b), there are two stability intervals close to the thin-wall limit and the thick-wall limit, respectively, in which the second excited $Q$-balls can resist the spherically symmetric perturbation.

Both stability intervals mentioned above are eliminated by the perturbation of non-spherical symmetry with $l=1$.
The mode $\lambda=1$, which behaves as an oscillation mode in the thin-wall region and transforms into an imaginary mode in the thick-wall region, makes the second excited $Q$-balls in the above stability interval near the thick-wall limit dynamically unstable.
On the other hand, the complex mode $\lambda=2$, which undergoes two transitions, triggers the dynamical instability of the second excited $Q$-balls in the other stability interval near the thin-wall limit.
At this point, one can conclude that the second excited $Q$-ball is vulnerable to a composite perturbation up to $l=1$.
In addition, the situation of the perturbation with $l=2$ is also presented, where the interaction between different mode pairs is observed for the first time.
As shown in Figure \ref{fig:29-34}(e), interestingly, the large oscillation mode in the first mode pair $\lambda=0$, which is migrating towards the interval boundary, will inevitably collide with the small oscillation mode in the second mode pair $\lambda=1$ migrating in the opposite direction, resulting in a transition that temporarily forms a pair of complex modes with conjugate symmetry. 

From the above results, it can be seen that the excited $Q$-balls with more nodes $n$ possess more unstable modes, making it more difficult to form a stability interval.
For the third excited $Q$-ball, even for the spherically symmetric perturbation, the existence of stability has not been observed.
%=======================================================================

%=======================================================================
\section{Conclusion and outlook}\label{sec:C}
In order to contribute to the theoretical framework for analyzing the dynamical stability of solitons, we introduce a numerical method based on the spectral decomposition technique, which has been widely used to study the stability of black holes \cite{Chung:2023zdq,Chung:2023wkd}.
The core idea of this method is to expand the background configuration and perturbation functions with a set of spectral basis functions that adaptively satisfy boundary conditions, and then reduce perturbation equations to an eigenvalue equation that can be easily solved numerically.
This method is not only simple but also universal.
Due to the self-interaction of solitons and the diversity of their geometric configurations, the use of traditional methods is greatly limited.
The spectral decomposition technique effectively overcomes these difficulties and is therefore expected to become a universal method to analyze the stability of solitons.

Adopting this method, the linear dynamical properties of $Q$-balls from the fundamental to the excited states are revealed systematically.
In the model presented in this work, the well-known stability criterion $dQ/d\omega<0$ holds true, indicating the stability of the fundamental $Q$-ball.
This is further confirmed by numerical results, where the fundamental mode satisfies $\Omega^{2}_{0}\geq 0$.
On the other hand, for the excited $Q$-balls, although the above stability criterion also holds, the system exhibits dynamical instability at both the spherically symmetric and non-spherically symmetric levels, manifested as the appearance of complex and imaginary modes.
In addition, the number of these unstable modes increases with the number of nodes in the configuration.

An important result of this work is the discovery of rich interactions between modes.
Through collision behavior, the oscillation mode representing stability and the imaginary or complex mode characterizing instability can be converted into each other, marking the transition in the dynamical properties of the system.
As a result, such a phenomenon leads to the existence of excited $Q$-balls capable of resisting perturbations with low-order spherical harmonics, indicating the rich dynamics of $Q$-balls, such as the dynamical transition between the fundamental $Q$-ball and the excited $Q$-balls.

From linear perturbation theory, this work reveals the dynamical properties of excited $Q$-balls near equilibrium, suggesting multiple evolution paths of the dynamics, depending on the geometry of the perturbation.
A natural direction in the future is to further investigate the real-time dynamics of excited $Q$-balls from nonlinear numerical simulations to verify these speculations.
In addition, the stability of solitons depends on the interaction of matter and the geometry of the configuration.
Therefore, an extension of this work is to explore the effects of the scalar potential and the charge repulsion (introducing a gauge field \cite{Lee:1988ag}) on the migration behavior of modes.
Another extension is to reveal the situation of solitons with different geometric configurations, such as spinning $Q$-balls \cite{Volkov:2002aj}, $Q$-shells \cite{Arodz:2008nm}, $Q$-strings \cite{Chen:2024axd}, $Q$-rings \cite{Axenides:2001pi}, $Q$-tubes and $Q$-crusts \cite{Sakai:2010ny}.

%=======================================================================

%=======================================================================
\section*{Acknowledgement}
We acknowledge the support from the China Postdoctoral Science Foundation with Grant No. 2024M760691 and the National Natural Science Foundation of China with Grants No. 12447129, No. W2431012, No. 12525503, and No. 12447101,
and the use of the Python packages Numpy \cite{Harris_2020}, Scipy \cite{Virtanen_2020} and Matplotlib \cite{4160265}.
%=======================================================================

\bibliographystyle{JHEP}
\bibliography{references}

\end{document}